\documentclass[%
 reprint,nofootinbib,superscriptaddress,
 amsmath,amssymb,widetext,
 aps,
pre,
]{revtex4-1}

\usepackage{ulem}
\usepackage{graphicx} 
\usepackage{bm}
\usepackage{txfonts}
\usepackage{mathrsfs}
\usepackage{amsmath}
\usepackage{xcolor}

\def\b0{{\mathbf 0}}

\def\b0{{\mathbf 0}}

\def\beq{\begin{equation}}
\def\eeq{\end{equation}}
\def\beqa\begin{eqnarray}
\def\eeqa{\end{eqnarray}}

\begin{document}

\title{The $q-$state Potts model from the Nonperturbative Renormalization Group}

\author{Carlos A. S\'anchez-Villalobos }
\affiliation{Sorbonne Universit\'e, CNRS, Laboratoire de Physique Th\'eorique de la Mati\`ere Condens\'ee, LPTMC, 75005 Paris, France}
\affiliation{Instituto de F\'isica, Facultad de Ingenier\'ia, Universidad de la Rep\'ublica, J.H.y Reissig 565, 11300 Montevideo, Uruguay}
\author{Bertrand Delamotte}
\affiliation{Sorbonne Universit\'e, CNRS, Laboratoire de Physique Th\'eorique de la Mati\`ere Condens\'ee, LPTMC, 75005 Paris, France}
\author{Nicol\'as Wschebor}
\affiliation{Instituto de F\'isica, Facultad de Ingenier\'ia, Universidad de la Rep\'ublica, J.H.y Reissig 565, 11300 Montevideo, Uruguay} 
\date{\today}
\begin{abstract}
We study the $q$-state Potts model for $q$ and the space dimension $d$ arbitrary real numbers  using the Derivative Expansion of the Nonperturbative Renormalization Group at its leading order, the local potential approximation (LPA and LPA'). We determine the curve $q_c(d)$ separating the first ($q>q_c(d)$) and second ($q<q_c(d)$) order phase transition regions for $2.8<d\leq 4$. At small $\epsilon=4-d$ and $\delta=q-2$ the calculation is performed in a double expansion in these parameters and we find $q_c(d)=2+a \epsilon^2$ with $a\simeq 0.1$. For finite values of $\epsilon$ and $\delta$, we obtain this curve by integrating the LPA and LPA' flow equations. We find that $q_c(d=3)=2.11(7)$ which confirms that the transition is of first order in $d=3$ for the three-state Potts model.
\end{abstract}


\maketitle

\section{Introduction}

Together with the clock models, the $q-$state Potts model \cite{Potts:1951rk,Wu:1982ra} is the most natural and famous generalization of the Ising model in terms of discrete degrees of freedom. It consists of lattice models where, at each site, a ``spin" can be in $q$ possible states and the Hamiltonian is symmetric under any permutation of these states. 

Beyond its academic interest, the $q-$state Potts model is physically relevant in different physical instances. For instance, for $q=3$  in dimension $d=3$, it describes the liquid crystal nematic-isotropic transition \cite{deGennes71}, a structural cubic to tetragonal crystal transition \cite{Weger19741}  as well as the confinement/deconfinement phase transition in pure Yang-Mills theory at finite temperature \cite{Polyakov:1978vu,Svetitsky1982423,Yaffe1982963}. In $d=2$, it describes the lattice gas transition of $^4$He atoms adsorbed on graphoil \cite{Alexander1975353,Bretz1977501,Tejwani1980152}. It has also been suggested that the 4-state model could be relevant to phase transitions in some antiferromagnets \cite{domany1982type}. The analytic continuation to $q=1$ and $q=0$ enables the study, respectively, of bond percolation and the spanning forest universality classes \cite{Fortuin:1971dw,Baxter:1973jmd,Harris1975327,lubensky1978field,Sokal_2005,BenAliZinati:2017vjy}.

From a theoretical point of view, the $q$-state Potts model is both a much-studied model for which many exact results are known in dimension $d=2$, and a model for which the physics in $d>2$ is poorly understood. For example, the mean field analysis \cite{Wu:1982ra,Kihara54,deGennes71,Straley_1973,Mittag_1974} predicts a first order transition in all dimensions for all $q>2$ which contradicts an exact result by Baxter showing that the transition is of second order in $d=2$ for $q\le4$ \cite{Baxter:1973jmd}. On the other hand, a simple dimensional argument suggests that for $q>2$, the upper critical dimension of the model is six, yet the $\epsilon=6-d$ expansion to the order of two loops \cite{Amit:1976pz,deAlcantaraBonfim:1980pe,deAlcantaraBonfim:1981sy}  leads to physically absurd results  with regard to the critical properties of the model.

The origin of these disturbing results is probably that for $q>2$, the Hamiltonian of the model involves a cubic term which on one hand allows for a systematic perturbative expansion in $\epsilon=6-d$ but, on the other hand, yields a thermodynamic potential which is unbounded from below. Notice that this $\epsilon$-expansion is under control for  $q=0$ and $q=1$ for which the instability of the potential is unlikely to be a problem. The scaling found in the $\epsilon$-expansion for $q>2$ probably corresponds only to scaling in a metastable state \cite{PriestLubensky,PriestLubenskyerratum} and not to a true second order transition. This is the signal that the critical physics of the Potts model for $q>2$ is particularly subtle and that nonperturbative methods are needed. It also explains why this subject has been almost abandoned for decades except for some isolated studies using approximate methods \cite{Nienhuis81,Barkema91,GrollauTarjus2001} and for the recent study based on the conformal bootstrap approach \cite{Chester:2022hzt}.

Most previous studies have focused mainly on the three-dimensional case and indicate that for $q=d=3$ the phase transition is of first-order
\cite{Nienhuis81,Barkema91,lee1991three,Alves91,janke1997three,GrollauTarjus2001}. Note that most of these results come from numerical simulations or studies of particular models, and are therefore only valid for those models. However, finding systems with $q=d=3$ that undergo a first order transition is not  proof that all systems undergo such a transition. It only means that, if any, they are outside the parameter region of second order transitions. The only way to decide whether or not a second-order transition is possible for $q=d=3$ is to prove or disprove that scale-invariance is possible in this case. This is what the renormalization group and the conformal bootstrap method allow for.

As for the conformal bootstrap, it has been extended to non-integer values of $d$ and suggests that for $q=3$ the dimension where the transition goes from second to first order is $d_c(q=3)\simeq 2.5$, so that for $q=3$, the transition is of first order in $d=3$. It should be noted that the unitarity-based bounds used in the conformal bootstrap approach are not rigorous for non-integer $d$ values. However, previous works suggest that these unitarity violations for non-integer $d$ have minor effects \cite{Hogervorst:2015akt,Chester:2022hzt}.

For what follows, it is important to note that a non-perturbative definition of the Potts model for arbitrary real values of $q$ exists \cite{Fortuin:1971dw} and the model can therefore be formulated for all real values of both $q$ and $d$.  It is therefore natural to try to determine in the $(d,q)$ plane, the $d_c(q)$ curve separating the small-$q$ second-order region from the large-$q$ first-order region. This curve cannot be obtained perturbatively because, as explained above, there is no value of $d$ where the perturbative expansion is under control. Our aim is to revisit this problem using the nonperturbative renormalization group (NPRG) which is a modern version of Wilson's  RG and to compute the  $d_c(q)$ curve at least down to $d=3$. This method has been used previously to study the $q-$state Potts model \cite{BenAliZinati:2017vjy}, but controlled results have so far only been obtained for $q=0$ and $q=1$. In the present work, we study the very different case $q\ge 2$.

Compared with Wilson's RG, the NPRG  shows several technical advantages when approximations are implemented which allows us to better control them (for a recent review of the NPRG and its applications, see \cite{Dupuis2021}). The derivative expansion (DE) is one of these approximation schemes and  we use it in the following. It has been widely used in the last twenty five years  with undeniable successes both in high energy physics and in statistical mechanics, at and out of equilibrium (see \cite{Dupuis2021} and Sec.~\ref{NPRG}). In recent years, it has been possible to explain the reason for all these successes, that previously remained rather obscure, by exhibiting a ``small parameter" associated with the DE \cite{Balog2019,DePolsi2020,DePolsi2021,DePolsi2022}. As a by-product, this also explains why this method is so versatile and robust. 

Despite what has been stated above and depending on the model, the implementation of the DE can be technically involved. This is precisely the case for the 3-state Potts model, which turns out to be a very difficult case for a variety of reasons, some of them technical and others related to the physics of the problem. The technical reasons are detailed later in this article, but there are some physical reasons that should be mentioned at the outset. First, as already mentioned,  there are no limits in which the $q=3$ case can be treated in a perturbative way. This makes it extremely difficult to test the quality of the approximations. Second, all the known results about the Potts model  suggest that  $d_c(q=3)<2.5$ \cite{Chester:2022hzt}. However, the leading order of the DE (usually dubbed the Local Potential Approximation, or LPA) has been tested as a function of $d$ for a great variety of models and is usually no longer reliable in low dimensions, typically $d\lesssim 2.5$ \cite{Chlebicki2022}. This means that the dimensions in which we expect to find a fixed point (FP) of the RG associated with a second-order phase transition are precisely those for which  the application of the LPA becomes doubtful. The LPA is therefore not an option to compute $d_c(q=3)$ which implies going directly to the next order. However, the second order of the DE is technically and numerically very difficult and goes far beyond this first study of $d_c(q)$.

Fortunately, as was observed a long time ago by Newman {\it et al.}, near $d=4$ and $q=2$, although the $q-$state Potts model  is not perturbative in the usual sense, a modified perturbative theory makes it possible to determine the shape of the curve $d_c(q)$ \cite{Newman:1984hy}. Indeed, for small $\delta=q-2$, the model is close to the Ising model which can be controlled by perturbation theory in $\epsilon=4-d$. This allowed these authors to prove under very mild assumptions that the curve $q_c(d)$ behaves near $d=4$ as $q_c(d=4-\epsilon)=2+a \epsilon^2$ with  $a$ a constant that is not determined by perturbation theory. This semi-perturbative regime is an ideal  starting point for implementing approximate but nonperturbative methods which is what we are doing below. 

In fact, the authors of Ref.~\cite{Newman:1984hy} have implemented a nonperturbative approximation to compute the curve $q_c(d)$ in the context of Wilson's RG. They have truncated the exact RG flow by projecting it onto a restricted space of coupling terms involving at most eleven couplings, that is, up to terms of the potential of order 6 in the fields. Unfortunately, this approximation is too short to achieve a converged determination of $q_c(d)$  below $d\sim 3.4$ and the most interesting case corresponding to $d=3$ has so far remained inaccessible. In the present work, we implement a similar scheme in the context of the NPRG and we include thirty couplings, that is, up to terms of the potential of order 9 in the fields. This allows us to reach the important $d=3$ case. We  obtain  $q_c(d=3)=2.11(7)$. Compared with previous works, our study of $q_c(d)$ has the double advantage of starting from a fully controlled point around $d=4$ and $q=2$ and being able to reach $d=3$.

The article is organized as follows. We present the $q-$state Potts model, its symmetries and the associated mean-field analysis in Sec.~\ref{Sec:model}. In Sec.~\ref{NPRG}, we present the NPRG method and the approximation scheme (the DE) that we implement at leading order in the present work. In Sec.~\ref{Invariants_and_tensors}, we build the tensors and scalars of the permutation group ${\cal S}_q$ relevant for the $q-$state Potts model and necessary to derive the RG flow equations. In Sec.~\ref{Secqc}, the main result of the article is presented: the curve $q_c(d)$. We conclude by a summary and an outlook. Some technical details are given in Appendices.

\section{The model and its main properties}
\label{Sec:model}
\subsection{Two equivalent lattice formulations of the Potts model}
\begin{figure}[t]
    \centering
    \includegraphics[scale=1.0]{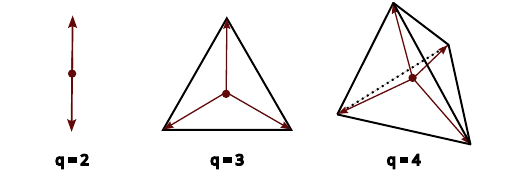}
    \caption{Vectors $\vec e^{\,(\alpha)}$ describing the possible values of $\vec S_i$ for $q=2, 3$ and $4$ in the Hamiltonian of Eq.~(\ref{Hamiltonian}).}
    \label{fig:vectors}
\end{figure}
The $q$-state Potts model is one of the simplest generalizations of the Ising model in which each spin can have $q$ possible states, all playing equivalent roles \cite{Potts:1951rk,Wu:1982ra}. As a result, the model is ${\cal S}_q$ symmetric, that is, is invariant under all permutations of the states. The simplest lattice Hamiltonian showing this symmetry is:

\begin{equation}
\label{HPotts}
 H=-J\sum_{\langle i,j\rangle} \delta_{\sigma_i,\sigma_j}
 \end{equation}
where the spins $\sigma_i \in \{1,2,\dots,q\}$, the sum is performed on nearest neighbor sites of a $d$-dimensional lattice and the model is ferromagnetic (anti-ferromagnetic) if $J>0$ ($J<0$). In the following, we are only interested  in the ferromagnetic Potts model.

The model can be written equivalently in terms of vector spins $\vec S_i$ with $n=q-1$ components. At each lattice site $i$, the spin $\vec S_i$ belongs to the set $\{\vec e^{\,(1)},\vec e^{\,(2)},\dots,\vec e^{\,(q)}\}$ where the $\vec e^{\,(\alpha)}$ are vectors joining the barycenter of a $n-$dimensional regular hyper-tetrahedron to its vertices, see Fig. \ref{fig:vectors}. The Hamiltonian is then:
\begin{equation}
\label{Hamiltonian}
 \tilde{H}=-\tilde{J}\sum_{\langle i,j\rangle} \vec S_i\cdot \vec S_j.
\end{equation}
and is therefore very similar to the ferromagnetic O($n$) model up to the difference that the spins $\vec S_i$ point in a discrete set of directions. 
The $q$ vectors $\vec e^{\,(\alpha)}$ are not independent since:
\begin{equation}
\label{baricenter}
\sum_{\alpha=1}^{q} \vec e^{\,(\alpha)}=0.
\end{equation}
They satisfy: 
\begin{equation}
\label{norm}
 \vec e^{\,(\alpha)}\cdot \vec e^{\,(\beta)}= A \delta_{\alpha\beta}+B
\end{equation}
because all vectors play a symmetric role.
This relation shows that the two Hamiltonians $H$ and $\tilde{H}$ are equivalent (up to an additive constant) under the condition that $J= A \tilde{J}$.

\subsection{The Ginzburg-Landau model}

In the field theoretical approach to critical phenomena, it is convenient to work with fields $\vec\varphi(x)$ that are unconstrained, that is, whose direction is not necessarily one of the $\vec e^{\,(\alpha)}$ and whose components vary between $-\infty$ and $+\infty$. A potential $U(\vec\phi)$ contributing to the Hamiltonian replaces the ``hard constraints" satisfied by the vectors $\vec S$  in Eq.~\eqref{Hamiltonian}  with ``soft constraints" that penalize configurations of the $\vec\phi$ different from those of the $\vec S$. The resulting Hamiltonian is called the Ginzburg-Landau (GL) Hamiltonian. It reads on the lattice:
\begin{equation}
H_{\rm GL}[\vec\varphi]=-\tilde{J}\sum_{\langle i,j\rangle} \vec \varphi_i\cdot \vec \varphi_j+\sum_i U(\vec\varphi_i)
\end{equation}
and, after rescalings, its continuum version is: 
\begin{equation}
\label{S_GL}
 H_{\rm GL}[\vec\varphi]=\int d^dx\,\Bigg( \frac 1 2 \big(\partial_\mu\vec \varphi(x)\big)^2+U(\vec\varphi(x))\Bigg).
\end{equation}
In $H_{\rm GL}$, the potential $U$ must have its $q$ minima pointing in the direction of the vertices of a $(q-1)$-dimensional tetrahedron. The problem of building $H_{\rm GL}$ thus boils down to that of building a general  ${\cal S}_q$-invariant potential $U$. 

Notice that in general the hard constraints satisfied by the $\vec S$ can be recovered on the $\vec\varphi$ in the limit where $\exp(-U(\vec\varphi))$ becomes a Dirac function that selects only the configurations of the $\vec S$. The original model, Eq. \eqref{Hamiltonian}, and the Ginzburg-Landau model are thus expected to be in the same universality class when they both undergo a continuous transition. In most cases, a truncation of $U(\vec\varphi)$ keeping only the nontrivial terms of lowest degree in the fields is sufficient to pick up one model belonging to the universality class. However, as we show below, the RG flow we are interested in couples all invariants and it is therefore mandatory to build all of them.

\subsection{ Scalars and tensors}
\label{scal-and-tens}

In the following, we need the construction of the invariant tensors and of the scalars of the model which requires the explicit construction of the vectors $\vec e^{\,(\alpha)}$. These different constructions have been done in the literature \cite{Golner73,Zia:1975ha,BenAliZinati:2017vjy} and we recall them below for the sake of completeness. Let us first show that the normalization of the vectors $\vec e^{\,(\alpha)}$ can be important from a practical point of view.

The constants $A$ and $B$ in Eq.~\eqref{norm} are not independent. Taking the square of the identity (\ref{baricenter}), one finds that for $\alpha\neq \beta$:
\begin{equation}
\vec e^{\,(\alpha)}\cdot \vec e^{\,(\beta)}=-\frac{1}{q-1}|\vec e^{\,(\alpha)}|^2
\end{equation}
and thus
\begin{equation}
\label{scalarprod}
 \vec e^{\,(\alpha)}\cdot \vec e^{\,(\beta)}=\frac{|\vec e^{\,(\alpha)}|^2}{q-1}
 \big(q\,\delta_{\alpha\beta}-1\big).
\end{equation}

Whenever the limit $n\to0$ has to be taken, it is convenient to choose the normalization: $|\vec{e}^{\,(\alpha)}|^2=q-1=n$ \cite{BenAliZinati:2017vjy}. Since we are interested in finite values of $n$, we choose:
\begin{equation}
\label{normcond}
|\vec{e}^{\,(\alpha)}| = \sqrt{\frac{2n}{n+1}}
\end{equation}
 from which follows
\begin{equation}
\label{scalarprodnorm}
 \vec e^{\,(\alpha)}\cdot \vec e^{\,(\beta)}=2
 \,\Big(\delta_{\alpha\beta}-\frac{1}{n+1}\Big).
\end{equation}

The general construction of the vectors $\vec{e}^{\,(\alpha)}$ is presented in Appendix~\ref{explicit vectors} together with some of their properties. 

We can now build the ${\cal S}_q$-invariants contributing to $U(\vec\phi)$, that is, invariants that do not include any derivative of the fields. As any permutation of $q$ objects can be decomposed into a succession of permutations between two objects, it is sufficient to require the invariance of the potential $U$ under all permutations $R^{(\alpha,\beta)}$ interchanging the vectors $\vec e^{\,(\alpha)}$ and $\vec e^{\,(\beta)}$ without modifying the others. 

A general polynomial in the coordinates $(\varphi_{i_1},\dots,\varphi_{i_n})$ of $\vec\varphi$ involving only terms of degree $p$ can be written:
\begin{equation}
\bar{T}^{(p)}_{i_1 i_2 \dots i_p}\varphi_{i_1}\dots\varphi_{i_p},
\end{equation}
where Einstein's convention is used, as will be done in the rest of the article. Without loss of generality, $\bar{T}^{(p)}_{i_1 i_2 \dots i_p}$ can be taken completely symmetric. The previous polynomial is invariant under ${\cal S}_q$ if and only if $\bar{T}^{(p)}$ is a completely symmetric invariant tensor of ${\cal S}_q$ of rank $p$. In this case, the transformation of the $\varphi_{i}$ under ${\cal S}_q$ is compensated by the invariance of $\bar{T}^{(p)}$ and the polynomial is invariant.

Obviously, all O$(n)$-invariant tensors are invariant under ${\cal S}_{q=n+1}$ since it is a subgroup of O($n$). For the O($n$) group, the invariant tensors are all linear combinations of tensor products of the Kronecker delta. We call them the isotropic tensors. For example, the completely symmetric tensors of order two and four are:
\begin{equation}
 \bar{T}^{(2)}_{ij}=\delta_{ij},\hspace{1cm}
 S^{(4)}_{ijkl}=\delta_{ij}\delta_{kl}+\delta_{ik}\delta_{jl}+\delta_{il}\delta_{jk}.
\end{equation}
Once contracted with the fields, they yield powers of the unique O$(n)$ invariant:
\begin{equation}
\label{rho}
 \rho=\frac 1 2 \big(\varphi_1^2+\varphi_2^2+\dots+\varphi_n^2\big)
\end{equation}
since
\begin{equation}
 \bar{T}^{(2)}_{ij}\varphi_i\varphi_j=2 \rho, \hspace{1cm}
 S^{(4)}_{ijkl}\varphi_i\varphi_j\varphi_k\varphi_l=12 \rho^2.
\end{equation}
For the ${\cal S}_q$ group, there are many other algebraically independent invariant tensors. We call them anisotropic tensors because their presence in a Hamiltonian is the signature of the explicit O($n$) symmetry breaking down to the ${\cal S}_q$ symmetry. We now show  how to build the simplest one which is of rank 3.
 
From the basis vectors $\vec e^{\,(\alpha)}$ it is easy to build a rank-3 completely symmetric tensor: 
\begin{equation}
\label{def-T3}
 \bar{T}^{(3)}_{ijk}=\frac 1 2\sum_{\alpha=1}^{q} e^{(\alpha)}_{i}e^{(\alpha)}_{j} e^{(\alpha)}_{k}.
\end{equation}
A tensor is invariant under ${\cal S}_q$ if by applying any permutation $R^{(\beta,\gamma)}$ it remains unchanged. This is obvious for $\bar{T}^{(3)}$ from its definition since $R^{(\beta,\gamma)}$ acts only on the two terms of the sum in Eq. \eqref{def-T3} where $\alpha=\beta$ and $\alpha=\gamma$ and exchange them. An invariant polynomial of order three is therefore:
\begin{equation}
\label{tau3}
 \bar\tau_3= \frac 1 2 \bar{T}^{(3)}_{ijk}\varphi_i\varphi_j\varphi_k.
\end{equation}
Let us notice that $\bar{T}^{(3)}\equiv 0$ for $q=2$ because in this case, $\vec e^{\,(1)}=-\vec e^{\,(2)}$ are one-component vectors and $\bar{T}^{(3)}_{111}=0$. This is expected since $q=2$ corresponds to the Ising model which is known to have $\rho$ as only invariant. We show in Appendix~\ref{tensorprops} that for any value of $q$, $\bar{T}^{(3)}$ must have an even number of indices equal to 1 to be non-zero.

For any $q\ge3$, $\bar{T}^{(3)}\neq0$. We compute in Appendix~\ref{explicitT}  two components of $\bar{T}^{(3)}$ for any $q\ge3$:
\begin{equation}
 \bar{T}^{(3)}_{112}=-\bar{T}^{(3)}_{222}=-\frac{1}{\sqrt{3}}
\end{equation}
 which implies that
\begin{equation}
\label{tau_3explicit}
 \bar\tau_3\Big|_{\varphi_3=\dots=\varphi_n=0}= \frac{1}{2}\Big(3 \bar{T}^{(3)}_{112}\varphi_1^2 \varphi_2+ \bar{T}^{(3)}_{222}\varphi_2^3\Big)=\frac{1}{2\sqrt{3}}\big(\varphi_2^3-3
 \varphi_1^2 \varphi_2\big)
\end{equation}
independently of the values of $n$. This expression contains all the terms of $\bar\tau_3$ when $q=3$ but, of course, for $q>3$ other terms including $\varphi_3, \varphi_4, \cdots$ contribute to $\bar\tau_3$. 
Whether the projection of $\bar\tau_3$ onto the $\big(\varphi_1,\varphi_2)$ plane is independent of $q$ depends crucially on the choice of normalization condition,  Eq.~(\ref{normcond}). With other normalizations of  $\vec e^{\,(\alpha)}$ or of $\bar{T}^{(3)}$ this projection may depend on $q$ via a multiplicative factor or through a permutation of indices (see \cite{BenAliZinati:2017vjy}, for example). 

Let us now generalize the construction above to general tensors. The obvious generalization of Eq. \eqref{def-T3} is:
\begin{equation}
\label{indeptensors}
 \bar{T}^{(p)}_{i_1 i_2\dots i_p}=\frac 1 2\sum_{\alpha=1}^{q} e^{(\alpha)}_{i_1}e^{(\alpha)}_{i_2}\dots e^{(\alpha)}_{i_p}.
\end{equation}
and the proof that it is invariant under ${\cal S}_q$ is identical: any permutation $R^{(\alpha,\beta)}$ leaves the sum in Eq. \eqref{indeptensors} unchanged because it only exchanges two of its terms. It follows that 
\begin{equation}
\bar{\tau}_p= \frac 1 2 \bar{T}^{(p)}_{i_1 i_2 \dots i_p}\varphi_{i_1}\dots\varphi_{i_p}.
\end{equation}
is invariant under ${\cal S}_q$. The explicit construction made in  Appendix~\ref{tensorprops} shows that for integer values of $q$ the tensors $\{\bar{T}^{(2)},\bar{T}^{(3)},\dots,\bar{T}^{(q)}\}$ are independent. It is important to notice that  this is also a complete set of independent tensors because one cannot construct, for any group, more than $n$ independent invariants out of a $n-$component vector. This implies that all higher order invariant terms are sum of products of $\bar{\tau}_p$ with $p\le q$.

The tensors $\bar{T}^{(p)}$ enjoy many algebraic properties  reviewed in Sect.~\ref{Invariants_and_tensors}.

\subsection{The mean-field approximation}
\label{ssec:mean-field}
We recall below that the Potts model undergoes a first order transition at mean field level for all values of $q>2$ \cite{Wu:1982ra,Kihara54,deGennes71,Straley_1973,Mittag_1974}. To show this, it is sufficient to consider the GL Hamiltonian in its continuum version:
\begin{equation}
\label{H_GL}
 H_{\rm GL}[\vec\varphi]=\int d^dx\,\Big( \frac 1 2 \big(\partial_\mu\vec \varphi(x))^2+r \rho(x)+\frac{v}{3!} \bar\tau_3(x)+\frac{u}{6}\rho^2(x)+\cdots\Big)
\end{equation}
and to show that the transition cannot be of second order. 

The spirit of the mean-field approximation is either to neglect all fluctuations or, at least, to neglect long wavelength fluctuations. In this approximation, the Gibbs free energy is a smooth function of $\vec\phi=\langle\vec\varphi\rangle$ that can be expanded as $H_{\rm GL}$ in Eq. \eqref{H_GL}, but with effective couplings. Therefore, at small magnetization, the free energy per unit volume evaluated for a constant field is:
\begin{equation}
\label{eqn:meanfieldGamma}
\frac 1 V\Gamma(\vec \phi,T)= r_{\rm eff}(T)\, \rho+\frac{v_{\rm eff}(T)}{3!} \bar\tau_3+\mathcal{O}(|\vec \phi|^4),
\end{equation}
where $V$ is the space volume, $r_{\rm eff}(T)$ and $v_{\rm eff}(T)$ are effective parameters depending smoothly on the temperature and $\rho$ and $\bar\tau_3$ are given by Eqs.~\eqref{rho} and \eqref{tau3} with $\vec\varphi$ replaced by $\vec\phi$. 
\begin{figure}
    \centering
    \includegraphics[scale=0.85]{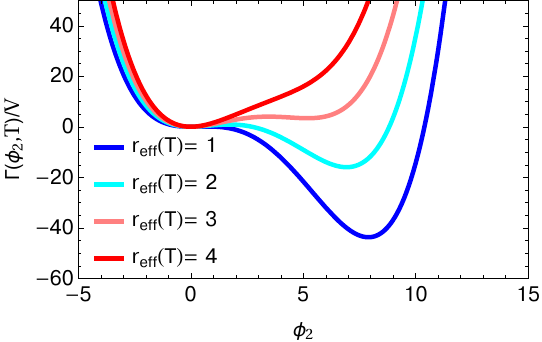}
    \caption{Free energy per unit volume at the mean-field approximation for $q=3$ and $\phi_1=0$. Units are chosen to make the magnetization and the free energy dimensionless. Three couplings have been retained: $r_{\rm eff}, v_{\rm eff}$ and $u_{\rm eff}$ which is the coupling of the $|\vec \phi|^4$ term. This term has been included to make the free energy bounded from below. At fixed $v_{\rm eff}=-10$ and $u_{\rm eff}=1$, the first order transition is induced by the variations of $r_{\rm eff}$ and occurs for $2<r_{\rm eff}< 3$.}
    \label{fig:meanf}
\end{figure}

If the transition were continuous, the magnetization would go to zero as $T$ goes to the transition temperature $T_c$ which requires $r_{\rm eff}(T_c)=0$. Now, if
$v_{\rm eff}(T_c)\neq 0$, the free energy does not have a minimum at $\vec \phi=0$ for $T=T_c$ because the trilinear term dominates at small fields and behaves for arbitrary values of $q$ as: 
\begin{equation}
 \frac{1}{V}\Gamma(\phi_1=\phi_3=\phi_4=\dots=0,T_c)=\frac{1}{2\sqrt{3}}\frac{v_{\rm eff}(T_c)}{3!} \phi_2^3+\mathcal{O}(\phi_2^4).
 \label{eq:mean-field-action}
\end{equation}
Therefore, the free-energy shows an inflexion point at zero field and not a minimum and the transition cannot be continuous, except if $v_{\rm eff}(T_c)=0$. However, this would not correspond to a critical point but to a tricritical point because two control parameters must be tuned at the same time to impose both $r_{\rm eff}(T_c)=0$ and $v_{\rm eff}(T_c)=0$. Therefore, at mean-field, the model cannot be critical  and if there is a transition corresponding to the tuning of one parameter only, it must be of first order. 

Of course, the expansion in Eq.~(\ref{eqn:meanfieldGamma}) does not yield a free energy bounded from below but this is nothing but the consequence of the Taylor expansion made at small fields. Including higher powers in the field such as a $|\vec \phi|^4$ term, the free energy can become stable. We show in Fig.~\ref{fig:meanf} how it typically deforms when $r_{\rm eff}$ is varied and how the first order transition occurs at mean field level.

\subsection{Critical behavior in $d=2$ and upper critical dimensions}
The Potts model model has been solved exactly in $d=2$ \cite{Baxter:1973jmd} and it has been proven that the transition is of first order for $q>4$ and continuous for $q\le 4$. This shows that, at least in low dimensions, the model shows important fluctuations that invalidate the mean-field analysis near the transition. Reciprocally, one can expect that for any $q$, the mean field approximation should be a reasonable approximation for large enough $d$. Usually, one defines the upper critical dimension $d_c$ of a given model as the dimension above which the universal critical properties are exactly taken into account by the mean field approximation. This implies that for $d>d_c$, the critical fluctuations are Gaussian. The scaling analysis is therefore performed around the Gaussian FP and is dominated by the couplings of largest engineering dimensions. For example, for the Ising model, the most relevant coupling is $u$ the scaling dimension of which is $4-d$. The critical dimension is thus $d_c=4$ and the critical theory is described  by the Gaussian FP for $d\ge4$. 

For $q>2$, the situation is very different. The most relevant interaction coupling with respect to the Gaussian FP is the trilinear one, whose Gaussian scaling dimension is $(6-d)/2$. As a consequence, if the transition were of second order the upper critical dimension of the model would be six, because above this dimension there are no relevant interactions at the Gaussian FP. A perturbative expansion in $\epsilon=6-d$ has been devised in \cite{Amit:1976pz,deAlcantaraBonfim:1980pe,deAlcantaraBonfim:1981sy}. Nevertheless, since the only relevant coupling gives rise to a potential that is not bounded from below, the results are not related to a critical transition, but to an expansion around a metastable state \cite{PriestLubensky,PriestLubenskyerratum}.

\begin{figure}
    \centering
    \includegraphics[scale=0.6]{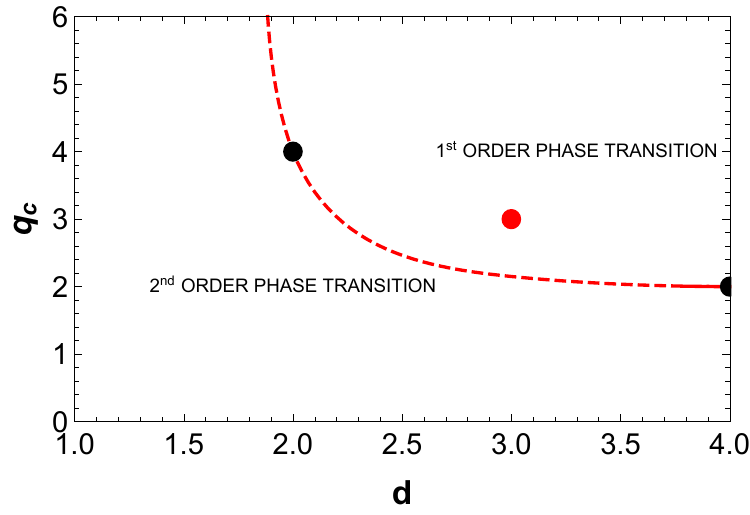}
    \caption{Known results for $q_c$ as a function of $d$. The dashed line corresponds to an interpolation using the known values, $q_c(d=2) = 4$ \cite{Baxter:1973jmd}, $q_c(d=4-\epsilon) = 2+\mathcal{O}(\epsilon^2)$ \cite{Newman:1984hy} and $q_c(1+\epsilon) \approx \exp(2/\epsilon)$ \cite{Andelman:1981yt,Nienhuis81}. The red point corresponds to $q=3$ and $d=3$, that is believed that is well inside the $1^{\rm st}$ order phase transition region \cite{Nienhuis81,Barkema91,lee1991three,Alves91,janke1997three,GrollauTarjus2001}.}
    \label{fig:qcvsdcartoon}
\end{figure}

Obviously, the potential needs to be bounded from below and this requires an even operator, that is, in the simplest case, a quartic term. For this term to be relevant, $d$ cannot be greater than 4 and we therefore expect that it is only below four dimensions that the transition can be of second order \footnote{This observation only applies when $q\in {\mathbb N}$ and it is unclear whether requiring  the potential to be bounded from below should apply to the analytic extension of the Potts model to noninteger values of $d$ or $q<2$. In fact, the perturbative expansion performed in $\epsilon=6-d$ seems to be under control for $q<2$ \cite{Gracey:2015tta,BenAliZinati:2017vjy} whereas it is not for $q>2$ \cite{PriestLubensky,PriestLubenskyerratum}}. A schematic representation of the curve that determines the boundary between a first-order and a second-order transition summarising the information known in the literature so far is shown in Fig. \ref{fig:qcvsdcartoon}.

A very interesting result was found by Newman {\it et al.} on this subject in Ref.~\cite{Newman:1984hy}. Using an extension of the model to non-integer values of both $q$ and $d$ these authors propose to study perturbatively the vicinity of the Ising model, that is, $q=2$ by making a double expansion in $\epsilon=4-d$ and $\delta=q-2$. They find a line $q_c(d)$ (or, equivalently, $d_c(q)$) below which the transition is of second order and above which it is of first order. It starts at $q=2$ and $d=4$ and $d_c(q)$ can be interpreted as the upper critical dimension for a given, generically noninteger, value of $q$  because for $d>d_c(q)$ the transition is of first order as predicted by the mean-field approximation. For a given $d$, they find two FPs for $q<q_c(d)$: a critical and a tricritical one. These two FPs collide  when $q=q_c(d)$ and become complex for $q>q_c(d)$: the transition becomes then of first order. Notice that this scenario of switching from a second to a first order transition is compatible with what is known in $d=2$ where at $q=4$, the critical  and tricritical FPs coincide and the transition becomes of first order for $q>4$ \cite{Baxter:1973jmd,Nienhuis:1979mb,Nienhuis_Riedel_Schick_1980}. 
It turns out that the calculation performed in \cite{Newman:1984hy} becomes unreliable for $d\lesssim 3.4$ and thus cannot address the case $q=3$ in $d=3$ which is the most important case. We discuss Newman's results in some detail in Sect.~\ref{neard4}. 

The calculation of $d_c(q=3)$ has been recently addressed with the Conformal Bootstrap approach in Ref. \cite{Chester:2022hzt}. The annihilation of the critical and  tricritical FPs is also observed and it has been shown that $d_c(q=3)\lesssim 2.5$. This results agrees with the common wisdom that for $q=d=3$  the transition is of first order \cite{Wu:1982ra}. Another interesting piece of information for the
$q_c(d)$ curve concerns the approach to  $d=1$ given by $q_c(1+\epsilon) \approx \exp(2/\epsilon)$ \cite{Andelman:1981yt,Nienhuis81}.

We are interested in the following in computing the line $q_c(d)$ defined above and in particular the value of $q_c(d=3)$. Since the perturbative method, that is, the $\epsilon=6-d$ expansion does not work for the Potts model we rely on a nonperturbative renormalization group method which is the modern version of Wilson's RG.

\section{The Nonperturbative Renormalization Group}
\label{NPRG}

In this section, we give a very brief overview of the Nonperturbative Renormalization Group (NPRG) and the approximation scheme that is used in this paper, the Local Potential Approximation, which is the leading order of the Derivative Expansion. Even if the content can be read in a much more detailed way in  many reviews (as \cite{Dupuis2021}), we include it for completeness.

\subsection{Nonperturbative Renormalization Group Equations}

The NPRG is based on Wilson's idea of integrating progressively short-distance degrees of freedom, that is, modes with a wavenumber larger than some scale $k$ while keeping the long-distance modes frozen. This is done by adding to the Hamiltonian (or euclidean action) of the model a quadratic term  that acts as an infrared regulator \cite{Polchinski:1983gv}, $H[\vec \varphi]\to H[\vec \varphi]+\Delta H_k[\vec \varphi]$ with:
\begin{equation} \label{deltaS}
\Delta H_k[\vec \varphi]=\frac 1 2 \int_{q}\varphi_i(-q) R_k(q^2) \varphi_i(q).
\end{equation}
Here and below $\int_q=\int \frac{d^dq}{(2\pi)^d}$. To act as a well-behaved infrared regulator, $R_k(q^2)$ must satisfy:
\begin{itemize}
	\item $R_k(q^2)$ is a $C^\infty$ function of the momentum squared;\footnote{This requirement can be relaxed in certain approximations.}
	\item $R_k(q^2)\sim Z_k k^2$ for $q\ll k$, where $Z_k$ is a field renormalization factor to be specified below;
	\item $R_k(q^2)\to 0$ very fast when $q\gg k$.
\end{itemize}

The infrared regularized free-energy $W_k[J]$ can be defined as usual \cite{Wetterich:1992yh,Ellwanger:1993kk,Morris:1993qb}:
\begin{equation} \label{regulatedgeneratingfunc}
e^{W_k[\vec J]}=\int \mathcal{D}\varphi\  e^{-S[\vec \varphi]-\Delta 
	H_k[\vec \varphi]+\int_x J_i(x) \varphi_i(x)}
\end{equation}
with $\int_x=\int d^dx$. Notice that the free energy $W[\vec J]$ of the original model is recovered in the limit $k\to0$ since $R_{k=0}\equiv 0$: $W_{k=0}[\vec J]=W[\vec J]$.

The regularized effective action $\Gamma_k[\vec \phi]$ is defined as a slightly modified Legendre transform of $W_k[J]$:
\begin{equation}
\label{legendre}
\Gamma_k[\vec \phi]=\int_x \phi_i(x) J^\phi_i(x) -W_k[\vec J^\phi]-\Delta S_k[\vec \phi]
\end{equation}
where $\vec J^\phi$ is a function of $\vec \phi$, determined implicitly by inverting the relation:
\begin{equation}
\phi_i(x)=\frac{\delta W_k}{\delta J_i(x)}\Bigg|_{J=J^\phi}.
\end{equation}
From the properties of the regulator $R_k(q^2)$ listed above and Eq.~(\ref{legendre}), it can then be shown \cite{Dupuis2021} that at a microscopic scale $k=\Lambda$, that must be much higher than any other dimensionful scale in the problem, $\Gamma_\Lambda[\vec \phi]\sim H[\vec \phi]$. This  provides the initial condition of the exact RG flow given below in Eq.~(\ref{wettericheq}).

The Gibbs free energy $\Gamma_k[\vec \phi]$ is the generating functional of infrared-regularized one-particle irreducible (1PI) proper vertices. In the following, we omit the $k$-dependence of the propagator and proper vertices to alleviate the notation. Once evaluated in a constant field $\vec\phi$, the Fourier transform of these vertices is defined by:
\begin{align} \label{eqgamma_n}
\Gamma^{(n)}_{i_1\dots i_n}(p_1,\dots,&p_{n-1};\vec \phi)=\int_x \,\mathrm{e}^{i \sum_{m=1}^{n-1}x_m\cdot 
	p_m} \nonumber \\
&\times\Gamma^{(n)}_{i_1\dots i_n}(x_1,\dots,x_{n-1},0;\vec \phi),
\end{align}
where
\begin{align}
\Gamma^{(n)}_{i_1\dots i_n}(x_1,\dots,&x_{n};\vec \phi)=\left.\frac{\delta^n \Gamma_k[\vec \phi]}{\delta \phi_{i_1}(x_1)\dots \delta \phi_{i_n}(x_n)}\right|_{\vec\phi(x)\equiv\vec\phi}.
\end{align}

The dependence of $\Gamma_k[\vec \phi]$ on $k$ or, equivalently, on the RG ``time" $t=\log(k/\Lambda)$ \cite{Wetterich:1992yh,Ellwanger:1993kk,Morris:1993qb} is given by:
\begin{equation}\label{wettericheq}
\partial_{t}\Gamma_{k}[\vec \phi]=\frac{1}{2}\int_{x,y}\partial_{t}R_{k}(x-y)G_{ii}[x,y;\vec \phi].
\end{equation}
Here $R_{k}(x-y)$ is the inverse Fourier transform of $R_{k}(q^2)$ and $G_{ij}[x,y;\vec \phi]$ is the full propagator in an arbitrary external field defined as the inverse of the two-point vertex function:
\begin{equation}
\int_{z} G_{il}[x,z;\vec \phi]\Big[\frac{\delta^2\Gamma_k[\vec \phi]}{\delta \phi_l(z)\delta \phi_j(y)}+ R_k(z-y)\delta_{lj}\Big]=\delta(x-y)\delta_{ij}.
\end{equation}
The scale-dependent effective potential is defined as the Gibbs free energy per unit volume evaluated in a constant field $\vec \phi$:
\begin{equation}
    U_k(\vec\phi)=\frac 1 V \Gamma_k(\vec\phi)
\end{equation}
It follows from Eq.~\eqref{wettericheq} that it satisfies an exact flow equation:
\begin{equation}
\label{eqpot}
\partial_t U_k(\vec \phi)=\frac 1 2 \int_q \partial_t R_k(q^2) G_{ii}(q,\vec \phi)
\end{equation}
where $G_{ij}(q,\vec \phi)$ is the Fourier transform of the propagator evaluated in the constant field $\vec \phi$. 

Equations for the $n$-point vertices in a constant external field can be obtained from Eq.~\eqref{wettericheq} by applying  $n$ functional derivatives. The flow equation of  $\Gamma^{(n)}$ is then expressed in terms of all the vertices up to $\Gamma^{(n+2)}$ which results in an infinite hierarchy of coupled NPRG equations. Solving this hierarchy requires approximations for most interacting theories (for a counter-example, see \cite{Benitez:2012sk}).

Compared with other common  approaches to field theory, the NPRG framework has the advantage of allowing approximations beyond perturbation theory. We now present the most widely used approximation in the context of NPRG, the derivative expansion.

\subsection{The derivative expansion and the issue of the infinite number of invariants}
The DE is an approximation scheme consisting in replacing $\Gamma_k[\vec\phi(x)]$ by its series expansion in the gradient of the field truncated at a finite order. For instance, for the Ising model, the DE truncated at its lowest order, called the Local Potential Approximation (LPA), consists in approximating $\Gamma_k[\phi(x)]$ by:
\begin{align}
\label{LPA-Ising}
& \Gamma_k^{\rm\, LPA}[\phi]= \int_x \Big(U_k(\phi)+ \frac{1}{2} 
 (\partial_\mu\phi)^2 \Big).
\end{align}
At fourth order of the DE and again for the Ising model it consists in approximating it by:
\begin{align}
\label{exDE}
& \Gamma_k^{\,\partial^4}[\phi]= \int_x \Big(U_k(\phi)+ \frac{1}{2} Z_k(\phi)
 (\partial_\mu\phi)^2\nonumber\\
&+\frac{1}{2} W^a_k(\phi)(\partial_\mu\partial_\nu\phi)^2 + 
\frac{1}{2} \phi W^b_k (\phi)(\partial^2\phi)(\partial_\mu\phi)^2 \nonumber\\
& + \frac{1}{2}  W^c_k(\phi)\left((\partial_\mu\phi)^2\right)^2
+\mathcal{O}(\partial^6) \Big).
\end{align}
To order $\partial^2$, it consists in keeping both $U_k$ and $Z_k$ and neglecting all other functions of the expansion. Due to the $\mathbb{Z}_2$ symmetry, the functions $U_k, Z_k, W^{a,b,c}_k, \cdots$ are functions of $\rho=\phi^2/2$ only. Once one of these  {\it ans{\"a}tze} is plugged into Eq. \eqref{wettericheq}  the NPRG equation boils down to a set of coupled partial differential equations on the functions involved in the {\it ansatz}. 

It has recently been demonstrated, both theoretically and empirically, that the DE is controlled by a small parameter for quantities defined at zero external momenta such as $U_k, Z_k, \dots$   from which can be computed thermodynamical quantities such as the correlation length, the magnetic susceptibility, the critical exponents, the universal equation of state, etc \cite{Balog2019}. Corrections to the leading order (LPA) are typically suppressed by a factor of the anomalous dimension $\eta$ \cite{Balog2019}. This makes the LPA a very well-suited approximation in cases where $\eta$ is small. Moreover, successive orders of the DE are suppressed by an expansion parameter which is rather small, of order $1/4$. Empirically the DE shows a rapidly converging behavior at least up to order $\partial^6$. This is to be contrasted with the usual perturbative expansions that are at best asymptotic Borel summable expansions requiring resummations techniques.

The convergence of the DE has been tested on the O($N$) models in $d=3$ by calculating critical exponents \cite{DePolsi2020,Peli:2020yiz} and universal amplitude ratios \cite{DePolsi2021}. In many cases, it gives the world best results for these quantities. The NPRG has also been used in many other contexts, such as disordered \cite{Tarjus:2004wyx,Tissier:2011zz} or non-equilibrium systems \cite{Canet:2003yu,Canet:2004je,Canet:2009vz}, almost systematically giving highly accurate results. It also makes it possible to calculate quantities that are beyond the reach of perturbative methods. Examples are nonuniversal quantities \cite{Parola95,Seide99,Canet:2004je,Machado10}  or FPs that are nonperturbative such as the strong coupling FP of the Kardar-Parisi-Zhang equation in $d=2$ and $d=3$ \cite{Canet:2009vz} or the breaking of supersymmetry in the random field Ising model in $d\simeq 5$ \cite{Tarjus:2004wyx,Tissier:2011zz}, see the review \cite{Dupuis2021} for an extensive bibliography. 

Our aim is to apply the DE at its leading order to the $q$-state Potts model. It is convenient to use both the LPA and a variant of the LPA, called the LPA', that consists in implementing a nontrivial field renormalization $Z_k$ on top of the LPA. The LPA' ansatz is
\begin{align}
\label{LPAPotts}
& \Gamma_k^{\rm LPA'}[\vec \phi]= \int d^dx \left(U_k(\vec \phi)+ \frac{1}{2} Z_k
 (\partial_\mu\vec \phi)^2\right)
\end{align}
where $Z_k$ is approximated by a field-independent quantity. The LPA is a simplification of the LPA' in which $Z_k$ is constrained to remain 1 all along the RG flow.

A difficulty specific to the $q$-state Potts model is that the number of invariants is $n=q-1$ whereas it remains equal to one for the O($N$) models independently of the value of $N$. The potential $U_k$ in Eq.~\eqref{LPAPotts} is therefore a function of $n$ variables, a much more complicated situation than for the O($N)$ models. 

As stated before, $q=2$ corresponds to the Ising universality class that has been largely studied in the literature, including high orders of the DE. The next integer value we are most interested in, presents a major difficulty: the value of $d_c(q=3)$ is expected to be around 2.5 \cite{Chester:2022hzt}. Below this dimension, the LPA is  expected to be a poor approximation because $\eta^{\rm LPA}=0$ whereas $\eta$ is probably not small for $d<2.5$. This is particularly visible in the behavior of the  effective potential at criticality, that is, $U_{k=0}(\vec \phi)$ at $T=T_c$, which is a power-law with exponent: $2d/(d-2+\eta)$. This power-law is clearly incompatible with $\eta=0$ in $d=2$ and the LPA is therefore invalid in this dimension.

The LPA' improves this situation since $\eta^{\rm LPA'}\neq0$ but this approximation is not under full control. The second order of the DE would be necessary to estimate the  confidence level of the LPA', but its implementation is challenging and then, for integer values of $q$ larger than two, a reliable analysis within the DE is very difficult.
 Moreover here, as in reference \cite{BenAliZinati:2017vjy}, we restrict ourselves to the simplest implementation of the DE that consists in performing a field expansion of $U_k(\Vec{\phi})$ on top of the LPA or LPA'.  A drawback of the field expansion of the LPA or LPA' is that it may fail to converge in small dimensions even if converges in $d=3$ \cite{Canet2003}. We study this in detail in the following. Our aim  is to show that we can push the expansion of $U_k(\Vec{\phi})$ to orders high enough for our results to converge in $d=3$.

A difficulty with the program described above comes from the continuation of $q$ to real values. Although all algebraic properties of the tensors defined in Sect. \ref{scal-and-tens} and \ref{Invariants_and_tensors} can be straightforwardly continued to real values of $q$, the RG flows for noninteger values of $q$ are tremendously more complicated than for integer values of $q$. The reason is simple: for  $q\in \mathbb{N}$, the number of independent invariants $\tau_p$ is finite and equals to $n=q-1$ which implies that the couplings associated with the invariants $\tau_{p>n}$ decouple from the flows of the other couplings with $p\le n$ [Note that this is only explicit if we choose a well-adapted parametrization of the tensors, which requires the definition of improved tensors, see Section \ref{Invariants_and_tensors}.]. For  $q\notin \mathbb{N}$, the above decoupling does not occur and it is necessary to keep in the {\it ansatz} for $\Gamma_k[\vec \phi]$ the infinity of invariants of the model whatever the order of the DE. At LPA for instance, the potential $U_k$ for a noninteger value of $q$ is a function of infinitely many invariants and it is therefore impossible to work functionally, even in principle. Fortunately, an expansion of $U_k$ in powers of the fields, similar to the expansion of the Hamiltonian in Eq. \eqref{H_GL}, does not show such a difficulty because a monomial of a given order in the fields involves only a finite number of invariants.

\section{Invariants, tensors and improved tensors}
\label{Invariants_and_tensors}
We show below that the NPRG flow of the coupling constants involved in the field expansion of $U_k$ requires the computation of contractions of several $\bar T^{(p)}$ tensors. These contractions are computed in Appendix \ref{tensorprops} and we review them below for completeness. We also show that since the number of independent tensors for a given $q\in {\mathbb N}$ is finite and equal to $n=q-1$, it is possible and convenient to build a set of improved tensors $T^{(p)}$ such that $T^{(p>q)}\equiv 0$ for any given $q\in {\mathbb N}$. The extension of these tensors to noninteger values of $n$ is also given below.

\subsection{Tensors and improved tensors}

As proven in Appendix~\ref{tensorprops}, the contraction of two tensors $\bar T$ is given by:
\begin{equation}
\label{contractionid}
\bar{T}^{(p)}_{i_1i_2\dots i_{p-1}k}\bar{T}^{(p')}_{j_1j_2\dots j_{p'-1}k}=\bar{T}^{(p+p'-2)}_{i_1i_2\dots i_{p-1}j_1j_2\dots j_{p'-1}}-\frac{2}{n+1}\bar{T}^{(p-1)}_{i_1i_2\dots i_{p-1}}\bar{T}^{(p'-1)}_{j_1j_2\dots j_{p'-1}}
\end{equation}
which implies for instance that:
\begin{equation}
\bar{T}^{(3)}_{ijm}\bar{T}^{(3)}_{klm}=\bar{T}^{(4)}_{ijkl}-\frac{2}{n+1}\delta_{ij}\delta_{kl}.
\end{equation}
Thus, the contraction of two tensors yields in general a higher rank  tensor. However, for $q\in\mathbb N$, the tensors with $p+p'-2>q$ in Eq. \eqref{contractionid} cannot be independent of the lower rank tensors and we show below that they are sums of products of these tensors.

Another useful property shown in Appendix~\ref{tensorprops} is:
\begin{equation}
\label{trace}
\bar{T}^{(p)}_{i_1i_2\dots i_{p-2}jj}=\frac{2n}{n+1}\bar{T}^{(p-2)}_{i_1i_2\dots i_{p-2}}
\end{equation}
An important feature of identities (\ref{contractionid}) and (\ref{trace}) is that they can be extended to non-integer values of $q$ as done in \cite{BenAliZinati:2017vjy}. We use this extension in Sec.~\ref{ssection:floweqLPA}.

As said above in Section \ref{scal-and-tens}, the $\bar T^{(p)}$ tensors can be divided into isotropic and anisotropic tensors depending on whether they are O($n$)-invariant tensors or not. However, this classification does not entirely fix what an anisotropic tensor is because to any of these tensors can be added an isotropic one while  remaining anisotropic. For instance, $\bar{T}^{(4)}$ can be modified by adding a multiple of $S^{(4)}$. It is therefore possible to modify the anisotropic tensors in such a way that they satisfy some extra properties. Following Ref.~\cite{Newman:1984hy}, we employ traceless tensors and more generally, we define the improved tensors $T^{(p)}$ by requiring that their full contraction with lower rank tensors is zero. For instance, for $p>2$, $T^{(p)}$ must satisfy: 
\begin{equation}
\label{partial-trace}
 T^{(p)}_{i_1i_2\dots i_{p-2}kl}T^{(2)}_{kl}=0,   
\end{equation}
that is, any partial trace must be zero. Notice that for $\bar{T}^{(3)}$:
\begin{equation}
 \bar{T}^{(3)}_{iij}=\frac 1 2\sum_{\alpha=1}^{q} e^{(\alpha)}_{i}e^{(\alpha)}_{i}e^{(\alpha)}_{j}=\frac{2 n}{n+1} \frac 1 2\sum_{\alpha=1}^{q} e^{(\alpha)}_{j}=0
\end{equation}
and thus $T^{(3)}=\bar{T}^{(3)}$. However this is no longer true for $\bar{T}^{(p>3)}$, as can be seen on Eq.~(\ref{trace}). The construction of $T^{(4)}$ is simple and we find that the traceless condition \eqref{partial-trace} imposes that
\begin{equation}
\label{improvedT}
T^{(4)}_{ijkl}=\bar{T}^{(4)}_{ijkl}-\frac{2n}{(n+1)(n+2)} S^{(4)}_{ijkl}.
\end{equation}
Notice that $T^{(4)}_{ijkl}T^{(3)}_{ijk}=0$ and $T^{(4)}$ is therefore the improved tensor of rank 4. 

It can be shown that for $p\le5$, the traceless condition \eqref{partial-trace} is sufficient to fully determine the improved tensors. That is, all other constraints coming from the contraction of $T^{(p\le5)}$ with $T^{(p'<p)}$ are automatically satisfied when Eq. \eqref{partial-trace} is. Starting from  $T^{(6)}$ this is no longer true and the contractions with $T^{(3)}, T^{(4)},\dots$ have to be taken into account to fully determine the improved tensors.

The $T^{(p)}$ defined above have many good properties. For example, we show in Appendix~\ref{tensorprops} that for $n=1$ or $n=2$, $T^{(4)}$ has, at most, only three nonvanishing components $T^{(4)}_{1111}$, $T^{(4)}_{1122}$ and $T^{(4)}_{2222}$ that are all proportional whatever the values of $n$: 
\begin{equation}
T^{(4)}_{1111}=T^{(4)}_{2222}=3 T^{(4)}_{1122}=\frac{(n-1)(n-2)}{(n+1)(n+2)}.
\end{equation}
This implies the important property that $T^{(4)}\equiv0$ for $n=1$ and $n=2$. More generally, for any given $q\in\mathbb N$, any improved tensor $T^{(p)}\equiv0$ for integer $p>q$.

\subsection{Invariants}
Once the tensors $T^{(p)}$ or $\bar T^{(p)}$ have been defined, the monomial in $\phi_i$ invariant under ${\cal S}_q$ can be constructed as in Eq. \eqref{tau3}. For instance, the improved invariants are:
\begin{equation}
\label{tau3-imp}
 \tau_p= \frac 1 2 T^{(p)}_{i_1i_2\cdots i_p}\phi_{i_1}\phi_{i_2}\cdots\phi_{i_p}.    
\end{equation}
The field expansion of the potential $U_k$ is the sum of the products of these invariants weighted by coupling constants, see Eq. \eqref{ansatzU} below for the expansion truncated to order 9 in powers of the fields. Notice that using either the $\tau_p$ or $\bar\tau_p$ invariants in this field expansion boils down to a linear redefinition of the couplings in front of them, which is immaterial  for the calculation of physical quantities. The only advantage of using improved invariants is to make manifest the fact that when $n\in\mathbb N$, only a finite set of invariants survive. This property is not explicit in the non-improved version and is not easy to check.

An interesting property of both improved and non improved invariants generalizes the one given for $\tau_3$ in Eq.~(\ref{tau_3explicit}). It is due to the normalization conventions that we have employed.  Consider two Potts models having $q$ and $q'$ states respectively, with  $q'>q$. Then the  invariants corresponding to the $q'-$state Potts model projected onto the space $\phi_{q}=\phi_{q+1}=\dots=\phi_{q'-1}=0$ are all identical to the  invariants of the $q-$state Potts model. This is trivial for the O($n$)-invariant $\rho$ defined in Eq. \eqref{rho}:
\begin{equation}
\label{projectionrho}
  \rho^{(q)}=\rho^{(q')}\big|_{\phi_{q}=\phi_{q+1}=\dots=\phi_{q'-1}=0}  
\end{equation}
and it can be shown to hold for all invariants:
\begin{equation}
\label{projectiontau}
  \tau_p^{(q)}=\tau_p^{(q')}\big|_{\phi_{q}=\phi_{q+1}=\dots=\phi_{q'-1}=0} .
\end{equation}

\subsection{Flow equations in the Local Potential Approximation}
\label{ssection:floweqLPA}
As said above, the study of the $q-$state Potts model for noninteger values of $q$ requires to perform a field expansion. For the LPA or LPA' defined in Eq.~(\ref{LPAPotts}), this amounts to expanding the potential $U_k(\vec\phi)\equiv U_k(\phi_1,\phi_2,\dots)$ in powers of the fields $\phi_i$. To order 9, the potential reads:
\begin{align}
\label{ansatzU}
U_k(\vec\phi)&=u_2 \rho+\frac{u_4}{6}\rho^2 +\frac{u_6}{90}\rho ^3+\frac{u_8}{2520}\rho^4
\nonumber\\
&+\frac{v_3}{3}\tau_3+\frac{v_5}{30}\rho \tau_3+\frac{v_6}{180}\tau_3^2+\frac{v_7}{630} \rho^2 \tau_3
   +\frac{v_8}{5040}\rho \tau_3^2 \nonumber\\
   &+ \frac{v_{9a}}{22680} \rho^3 \tau_3 +\frac{v_{9b}}{45360} \tau_3^3\nonumber+\frac{w_4}{12}\tau_4+\frac{w_6}{180}\rho \tau_4\\
&+\frac{w_7}{1260}\tau_3 \tau_4+\frac{w_{8a}}{5040}\rho^2 \tau_4+\frac{w_{8b}}{10080}\tau_4^2\\ &
+\frac{w_9}{45360} \rho \tau_3 \tau_4+\frac{x_5}{60}\tau_5+\frac{x_7}{1260}\rho \tau_5+\frac{x_8}{10080} \tau_3 \tau_5  \nonumber\\
&+ \frac{x_{9a}}{45360}\rho^2 \tau_5+\frac{x_{9b}}{90720} \tau_4 \tau_5+\frac{y_6}{360}\tau_6+\frac{y_8}{10080}\rho \tau_6\nonumber\\
&+\frac{y_9}{90720}\tau_3\tau_6+\frac{z_7}{2520}\tau_7+\frac{z_9}{90720} \rho \tau_7+\frac{s_8}{20160}\tau_8\nonumber\\
&+\frac{c_9}{181440} \tau_9\nonumber
\end{align}
where the  $u_a$'s are for terms  involving only $\rho$ and the index $a$ is the power of the fields, the $v_a$'s for terms  involving $\tau_3$ and $\rho$, the $w_a$'s for terms  involving $\tau_4$, $\tau_3$ and $\rho$, and so on. Notice that there are two terms of order 8 involving $\tau_4$ and we have called them $w_{8a}$ and $w_{8b}$. Below, as usual, we call $r= u_2$ and $u= u_4$.

The flow of $U_k(\vec\phi)$, given in Eq. \eqref{eqpot}, requires the computation of the full propagator, that is, of  $\big(\Gamma^{(2)}_k(q,\vec\phi)+R_k(q)\big)^{-1}$. In the LPA or LPA', $\Gamma^{(2)}_k(q,\vec\phi)$ is computed from Eqs. \eqref{LPAPotts} and \eqref{ansatzU}:
\begin{equation}
\Gamma_{ij}^{(2)}(q;\vec\phi)=Z_k q^2\delta_{ij}+\frac{\partial^2 U_k(\vec\phi)}{\partial\phi_i\partial\phi_j}=\big(Z_k q^2+r)\delta_{ij}+\frac{\partial^2 U_k^{\rm int}(\vec\phi)}{\partial\phi_i\partial\phi_j}
\end{equation}
where we have separated in $U_k$ the quadratic term  and  the interaction part  $U_k^{\rm int}(\vec\phi)$. The latter  includes at least a cubic term and its second derivative involves at least one field. Thus, expanding the propagator in powers of the field is equivalent to expanding in powers of 
\begin{equation}
    U_{ij}^{(2),{\rm int}}(\vec\phi)=\frac{\partial^2 U_k^{\rm int}(\vec\phi)}{\partial\phi_i\partial\phi_j}.
\end{equation}
Defining the propagator at zero field by:
\begin{equation}
G_{ij}(q)=\frac{\delta_{ij}}{Z_k q^2+R_k(q^2)+r}\equiv \delta_{ij}G(q),
\end{equation}
the expansion in powers of the field of the LPA flow equation of $U_k$ consists in inserting in Eq. \eqref{eqpot} the expansion
\begin{align}
G_{ij}(q;\vec\phi)=&\delta_{ij} G(q)-G^2(q)U_{ij}^{(2),\rm int}
+G^3(q)U_{il}^{(2),\rm int}U_{lj}^{(2),\rm int}
\nonumber\\
&-G^4(q)U_{il}^{(2),\rm int}U_{lm}^{(2),\rm int}U_{mj}^{(2),\rm int}+\cdots .
\end{align}
The flow equations of all the coupling constants can then be obtained from the flow of $U_k$ by projection onto the invariants $\tau_p$ or $\bar\tau_p$, depending on whether we want to work with ordinary or improved invariants. This requires tensor contractions which are straightforward using Eqs. (\ref{contractionid})  and (\ref{trace}) but which become increasingly tedious as the order of the truncation increases. Notice that the tensor contractions are simple for the nonimproved tensors and are more involved with improved tensors. Thus, for practical purpose, it is simpler to first work with nonimproved tensors and only at the end of the calculation to switch to improved couplings, if necessary.

Before discussing the flow equations of the coupling constants involved in Eq. \eqref{ansatzU}, let us define their dimensionless and renormalized counterparts. They are defined by \cite{Dupuis2021}:
\begin{align}
 U_k(\vec \phi)&=4 \omega_d\, k^d  \tilde{U}_k(\tilde{\vec\phi})\nonumber\\
\vec \phi&=2\sqrt{\omega_d}\, Z_k^{-1/2}k^{(d-2)/2} \tilde{\vec\phi}
\end{align}
where $\omega_d=\big(2^{d}\pi^{d/2} \Gamma(d/2)d\big)^{-1}$ and  $\eta_k=-\partial_t \log Z_k$ is the running anomalous dimension. A rescaling by the factor $\omega_d$ has been implemented so as to cancel
 large numbers coming from angular integration. The corresponding equation for the dimensionless potential is:
\begin{align}
\label{eqUtilde}
\partial_t \tilde{U}_k(\tilde{\vec\phi})
&+d \tilde{U}_k(\tilde{\vec\phi})-\frac{d-2+\eta_k}{2}\tilde{\phi}_i \frac{\partial \tilde{U}_k(\tilde{\vec\phi})}{\partial \tilde{\phi}_i}\nonumber\\
&=\frac{k^{-d}}{2\times 4 \omega_d}\int_q \partial_t R_k(q^2) G_{ii}(q,\vec\phi)
\end{align}
The field expansion of $\tilde{U}_k(\tilde{\vec\phi})$ to order 9 is similar to the one of Eq.~\eqref{ansatzU} with the invariants $\rho, \tau_3, \tau_4, \dots$ and the coupling constants $u_a, v_a, w_a, \dots$ replaced by their dimensionless counterparts. By collectively calling $g_m$ a coupling constant of a term involving $m$ fields, its dimensionless counterpart $\tilde g_m$ is given by:
\begin{equation}
g_m=\tilde{g}_m k^{d-m (d-2)/2}Z_k^{m/2}(4 \omega_d)^{(2-n)/2}.
\end{equation}
The flows of the dimensionless couplings $\tilde u_a, \tilde v_a, \tilde w_a, \dots$ are obtained by expanding both sides of Eq.~(\ref{eqUtilde}) in powers of the invariants. They only involve  the integrals
\begin{equation}
 I_n(r)=\int \frac{d^dq}{(2\pi)^d}\frac{\partial_t R_k(q^2)}{\big(Z_k q^2+R_k(q^2)+r\big)^n}
\end{equation}
or their dimensionless counterparts defined by
\begin{equation}
\label{dimensionlessIn}
 I_n(r)=\frac{4 \omega_d k^{d+2-2n}}{Z_k^{n-1}}\tilde{I}_n(\tilde r).
\end{equation}

From now on, we work only with dimensionless couplings and integrals and omit the tildes for simplicity.

For the sake of concreteness and because we need them in the following,
we give below the flow equations for the couplings corresponding to four fields or less. The other ones are given in a supplementary
material. These equations are:
\begin{align}
\label{flowr}
 \partial_t r=(\eta -2)r-\frac{(n+2)}{6}u I_2 +\frac{2(n-1)}{n+1}v_3^2  I_3 
\end{align}

\begin{align}
\label{flowv3}
\partial_t v_3&=\frac{1}{2} (d+3 \eta -6)v_3+\Bigg[2u+\frac{6 (n-2)}{n+2} w_4 \Bigg]v_3 I_3 \nonumber\\
&-\frac{n+6}{20}  v_5 I_2 -\frac{6(n-2)}{n+1}v_3^3I_4
\end{align}

\begin{align}
\label{flowu}
\partial_t u&=(d+2 \eta -4)u-\Bigg[\frac{n+4}{10} u_6+\frac{6 (n-1)}{5(1+n)(2+n)}v_6\Bigg]I_2
\nonumber\\
&+\Bigg[\frac{n+8}{3}  u^2+\frac{12 (n-1) (n+6)}{5 (n+1)
   (n+2)}v_3 v_5\nonumber\\
&+\frac{24 (n-2) (n-1)}{(n+1) (n+2)^2}w_4^2\Bigg]I_3
   \nonumber\\
&-\frac{12 (n-1) \Big(\big(n+2\big)\big(n+6\big)
   u_4+12 (n-2) w_4\Big)}{(n+1) (n+2)^2} v_3^2 I_4\nonumber\\
&+\frac{48
   (n-1) (3 n-4)}{(n+1)^2 (n+2)}v_3^4  I_5
\end{align}
\begin{align}
\label{floww4}
\partial_t w_4&=
(d+2 \eta -4)w_4 + \Bigg[v_3 \bigg(\frac{8 (n-3) (n+2) }{(n+1)
   (n+6)}x_5+\frac{12}{5}v_5\bigg)\nonumber\\
   &+6 \frac{n^2-3n-2}{(1+n)(2+n)} w_4^2+4 u w_4\Bigg] I_3\nonumber\\
&+ 12\Bigg[
   3\bigg(\frac{2 n + 4 - n^2 }{(n+1)(n+2)}\bigg) w_4-u\Bigg] v_3^2 I_4
   \nonumber\\
&-\frac{1}{30}
   \big((n+8) w_6+9 v_6\big) I_2 +\frac{24(n-3)}{n+1}v_3^4 I_5.
\end{align}
These flow equations must be completed by the expression of  $\eta_k$. At LPA, $\eta_k=0$ since $Z_k=1$ for all $k$. At LPA', the value of $\eta_k$ depends on the value of the field where it is computed. We choose here $\vec\phi=\vec 0$. Then, the value of $\eta_k$ is obtained from the flow equation of $\Gamma_{ij}^{(2)}(p^2;\vec\phi)$ expanded at order $p^2$ and evaluated  at $\vec\phi=\vec 0$. It is given by:
\begin{align}
\label{etaeq}
\eta_k &=2 \frac{n-1}{n+1} v_3^2 I_{\eta}.
\end{align}
Eq.~(\ref{etaeq}) involves a new dimensionless integral:
\begin{align}
I_{\eta}&=\frac{Z_k^2 k^{6-d}}{4 \omega_d}\int_q \partial_t R_k(q^2) G^4(q)
\Big\{Z_k+R_k'(q^2)\nonumber\\
&+2 \frac{q^2}{d}\Big[R_k''(q^2)-2 G(q) \big(Z_k+R_k'(q^2)\big)^2\Big]\Big\}.
 \end{align}
After some redefinitions of the couplings, we have checked that our equations coincide at order $\phi^6$ with those of Ref.~\cite{BenAliZinati:2017vjy} except for some typos in this reference, confirmed by the authors. 

A nice property of our flow equations is manifest on Eqs. 
(\ref{flowr}) to (\ref{floww4}). When $n=1$,  the flows of $r=u_2$ and $u=u_4$ no longer depend on $v_3, v_5$ or $w_4$ and when $n=2$, the flows of $r,u$ and $v_3$ no longer depend on $w_4$. One can check that this is a general phenomenon, independent of the LPA': when $n=1$ (Ising model)  the flows of all the $u$'s are independent of the $v$'s, $w$'s, $x$'s, etc; when $n=2$ the flows of all the $u$'s and $v$'s are independent of the $w$'s, $x$'s, etc, and so on. This property is on one hand trivial because for $n=2$ for instance, the potential depends only on $\rho$ and $\tau_3$ and their flow equations cannot involve other couplings. On the other hand, this property , that is independent of the choice of tensors, is manifest on the RG flow equations only with improved tensors, which is the advantage of working with these tensors.

It is important to realize that the argument above does not imply that for $n=1$ for instance, the couplings $v$'s, $w$'s, $x$'s, etc, vanish. They only decouple from the flows of the $u$'s in the limit $n\to1$ because in these flows they always contribute together with a prefactor proportional to $n-1$.

In the following, we use the $\Theta-$regulator \cite{Litim:2001up}:
\begin{equation}
\label{reg-theta}
 R^{\theta}_k(q)=Z_k (k^2-q^2) \Theta \big(1-q^2/k^2\big).
\end{equation}
This regulator is particularly convenient because it allows us to analytically compute the integrals:
\begin{align}
\label{thetaregIn}
 I_n&=\left(1-\frac{\eta}{d+2}\right)\frac{1}{(1+r)^n}  \nonumber\\
I_\eta&=\frac{1}{2} \frac{1}{(1+r)^4} .
\end{align}
Moreover, it has been shown empirically on the O($N$) models that at the LPA order, the best critical exponents are obtained with this regulator \cite{Litim:2001up,Canet2003} and although it does not regularize the DE from order $\partial^4$, it is optimal in this sense at LPA.

\section{The critical behavior of the $q-$state Potts model and the line $d_c(q)$}
\label{Secqc}

With the flow equations obtained to order $\phi^9$, we can study the existence and stability of the FPs of the $q-$state Potts model for continuous values of both $d$ and $q$. At order 9 of the field expansion of the LPA and LPA' flow equations, we calculate numerically the shape of the $q_c(d)$ curve separating in the $(d,q)$ plane the region of first-order phase transition that lies above this curve and the second-order region that lies below it, see Fig~\ref{fig:qcvsdcartoon}.
We show that  below this curve coexist a critical and a tricritical FP that collide when $q\to q_c(d)^-$ and disappear -- more precisely become complex -- for  $q> q_c(d)$.

Before doing that, following Ref.~\cite{Newman:1984hy}, we start this study by first analyzing the FPs existing in the neighborhood of the $q=2$ and $d=4$ that, in many aspects, can be solved exactly, that is, without requiring the LPA or LPA' approximations.

\subsection{Fixed points in $d=4-\epsilon$ and $q=2+\delta$}
\label{neard4}

The critical behavior of the Ising model is associated with the Wilson-Fisher FP  in $d<4$. When $d\to 4^-$, the Wilson-Fisher FP approaches the Gaussian FP which is tricritical and both FP collide in $d=4$ which is therefore the upper critical dimension of this model. We can therefore expect that $q_c(d=4)=2$. We should of course retrieve this from our flow equations except for one subtlety: when embedded in a set of more general models, here the $q-$state Potts models, a critical FP can become multicritical because the other couplings can be relevant at this FP. We therefore have to restudy the stability of the Ising FP, in particular when $n$ is close to 1 and $d$ close to 4 to determine $q_c(d)$ in the vicinity of $d=4$. This requires to determine the set of all FPs and we start by the perturbative ones, that is, those are close to the Gaussian FP when $d\to4$. It is interesting to note that perturbative multicritical FPs for $q=0$ and $q=1$ have been found in a perturbative analysis performed in $d=10/3-\epsilon$ \cite{Codello:2020mnt}.

\subsubsection{Perturbative fixed points in $d=4-\epsilon$}
\label{sssection:perturbFP}

\begin{figure*}
  \centering
\includegraphics[width=70mm]{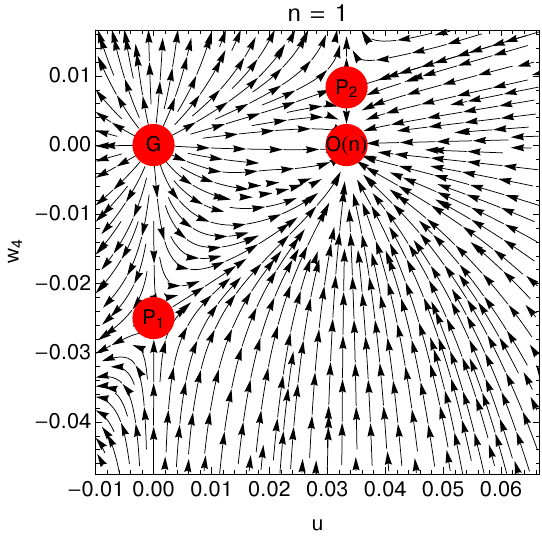}
\includegraphics[width=70mm]{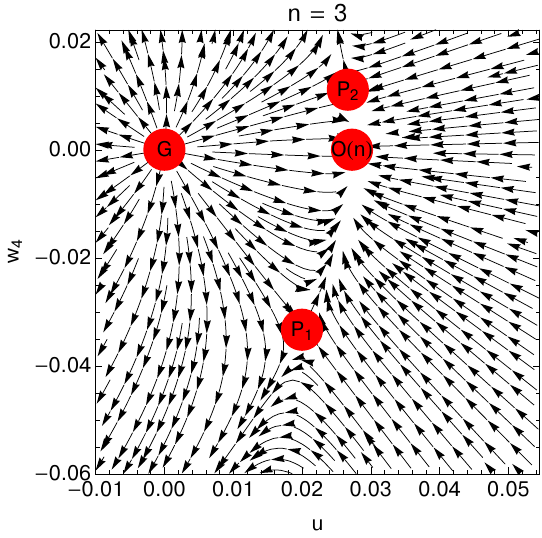}
\includegraphics[width=70mm]{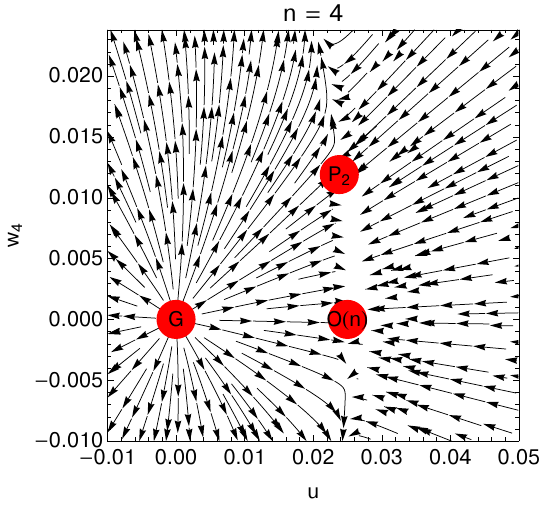}
\includegraphics[width=70mm]{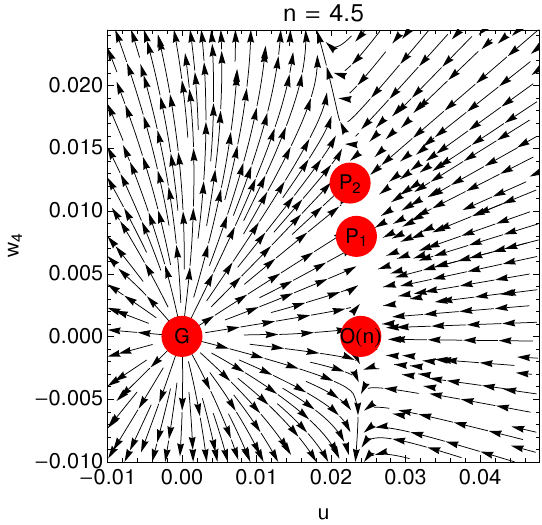}
\includegraphics[width=70mm]{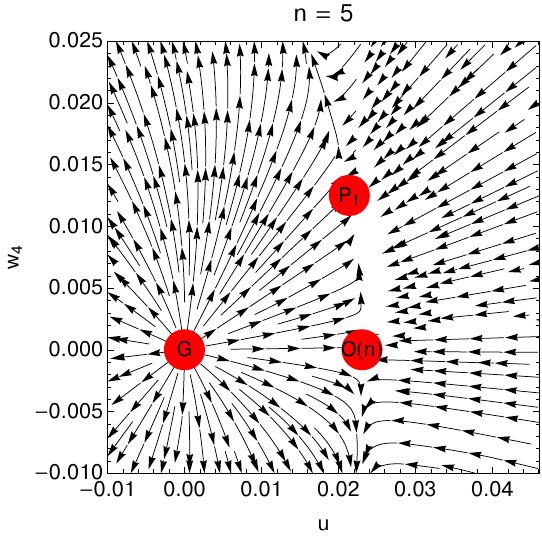}
\includegraphics[width=70mm]{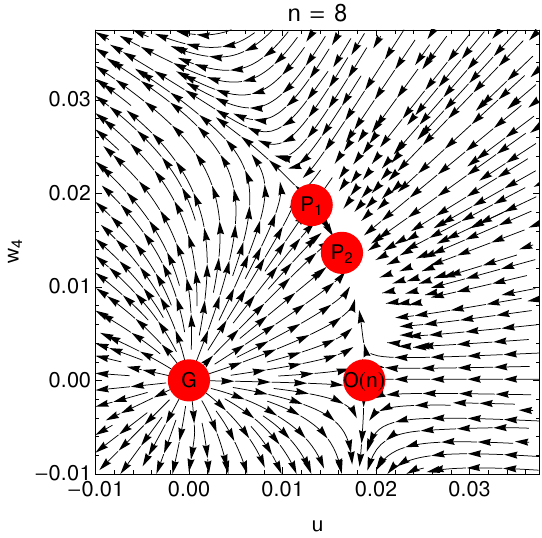}   
    \caption{Renormalization group flow in $d=3.9$ in the plane $(u,w_4)$ for $n  \in \{1,3,4,4.5,5,8\}$ and $k=4$. For the sake of simplicity, we have imposed $r=0$ and $v_3=0$. In the flow for $n=4$, the $O(n)-$invariant and P$_1$ FPs coincide. In the flow for $n=5$, the P$_1$ and P$_2$ FPs coincide. }
    \label{fig:flowperturd39lown}
\end{figure*}

As well-known \cite{Dupuis2021}, the LPA' flow of the potential is one-loop exact when $d\to 4$ and the couplings that are irrelevant with respect to the Gaussian FP can be neglected at  leading order in $\epsilon=4-d$. The relevance of the couplings near the Gaussian FP is given by dimensional analysis. By inspection, we find that only $r$, $v_3$, $u$ and $w_4$ with respective dimensions $2$, $1+\epsilon$, $\epsilon$ and $\epsilon$ are relevant with respect to the Gaussian FP in $d=4-\epsilon$. This FP is therefore pentacritical. All other perturbative FPs are at least tricritical because the scaling dimensions of $r$ and $v_3$ cannot become negative when moving from the Gaussian FP to a perturbative  FP which, by definition, is at a distance of order $\epsilon$ from the Gaussian FP. As a consequence, no perturbative FP can control a second order phase transition near $d=4$.

Neglecting all couplings associated with terms of degree higher than 4, we find that the FP equation for $v_3$ is given by:
\begin{equation}
v_3^*\left(\frac{1}{2}+\Big(2u^*+\frac{6 (n-2)}{n+2} w_4^* \Big)I_3
-\frac{6(n-2)}{n+1}(v_3^*)^2I_4\right)=0.
\end{equation}
Its only solution with couplings at most of order $\epsilon$ is $v_3^*=0$.  Notice that around the Gaussian FP, $v_3$ is the only relevant coupling constant associated with an odd term in the fields. Therefore, all perturbative FPs have  an extra $\mathbb{Z}_2$ symmetry consisting in changing all fields in their opposite. The total symmetry group is therefore enlarged to  $S_q\times \mathbb{Z}_2$ at these FPs.

We have found three Non-Gaussian FPs with couplings of order $\epsilon$. The first one is the usual O($n$)-invariant Wilson-Fisher FP with $v_3=w_4=0$ and $u\neq 0$. Notice that it is only for $n=1$ that this FP is in the Ising universality class: for a generic noninteger value of $n$ it is the extension to real values of $n$ of the Wilson-Fisher FP. Two other FPs with $w_4\neq0$ exist and are of order $\epsilon$. We call them P$_1$ and P$_2$. Their stability depends on the value of $n$. In all cases, there is one tricritical FP and two tetracritical ones. The tricritical FP is the Wilson-Fisher FP  for $n\le 4$,  P$_1$ for $4< n\le 5$ and  P$_2$ for $n>5$.
In $d=3.9$, the flows in the coupling constant space ($u,w_4$) are shown in Fig.~\ref{fig:flowperturd39lown} for various values of $n$. It is shown how the P$_1$ and $O(n)$ FPs exchange their stability at $n=4$ and the same for P$_1$ and P$_2$ at $n=5$.\footnote{To avoid the most relevant direction, we have replaced $r$ by zero in the flow equations to represent this figure. This does not modify the leading behavior of the flow equations for $u$ and $w_4$ near the FPs for $d\simeq4$.}

We conclude from the above discussion that if there exists a second order transition it cannot be controlled by a purely perturbative FP. It is shown in the next section that near $d=4$ and $n=1$ one can prove by a double expansion in $\epsilon=4-d$ and $\delta=n-1$ the existence of a critical FP. Although, as just discussed, this FP is not fully perturbative, several of its properties can be analysed perturbatively, so we will refer to it as ``semi-perturbative".

\subsubsection{Semi-perturbative fixed points in $d=4-\epsilon$ and $q=2+\delta$}
\label{sssection:semipert}

As we have seen above in Eqs.~(\ref{flowr}) to (\ref{floww4}), the flows of the couplings of the Ising model are recovered from the general flow equations of the $q-$state Potts model because in the $n\to1$ limit, the $v, w, x, \dots$ couplings decouple from the flows of the $u$'s as they are always accompanied by a factor $n-1$. This is  also the case for all isotropic couplings,  not only those included in the LPA', see Sec.~\ref{ssection:floweqLPA}. As shown by Newman, this is sufficient to derive a double expansion in  $\epsilon=4-d$ and $\delta=n-1$ for all FPs, including the critical one, where the isotropic sector can be analyzed perturbatively, except for some non-perturbative constants that come from the anisotropic sector, see below. 

We show now that in this double expansion the flows of the O($n$)-invariant couplings, that is, of the $u$'s, become perturbative without having recourse to the LPA or LPA'. In the limit $\delta\to0$, they become the flows of the couplings of the Ising model in $d=4-\epsilon$ and for  $\delta$  nonvanishing and small these flows are modified by terms of order $\delta$. For $n=1$, the flow of $u$ reads:
\begin{equation}
\label{perturbflow}
 \partial_t u=-\epsilon u+ 3 u^2 I_3+\mathcal{O}\big(\epsilon^3\big)
\end{equation}
where $I_3$ is 
\begin{equation}
 I_3  =1+\mathcal{O}\left(\epsilon\right)
\end{equation}
independently of the choice of regulator $R_k(q)$. Eq. \eqref{perturbflow} follows from the scaling in $\epsilon$ of the couplings near the Wilson-Fisher FP: $r, u\sim \mathcal{O}(\epsilon)$, the other isotropic couplings such as $u_6$ are, at least, of order $\epsilon^3$, $\eta\sim\mathcal{O}(\epsilon)^2$ and $\Gamma^{(4)}\sim u +\mathcal{O}(\epsilon)^2$.

\begin{figure}[t!]
    \centering
    \includegraphics[scale=0.65]{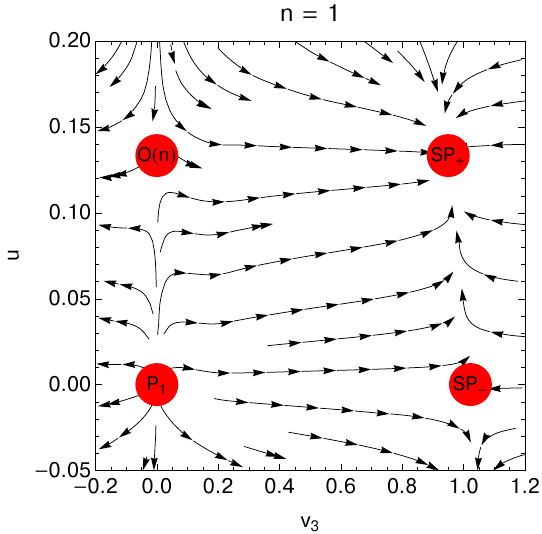}
    \includegraphics[scale=0.65]{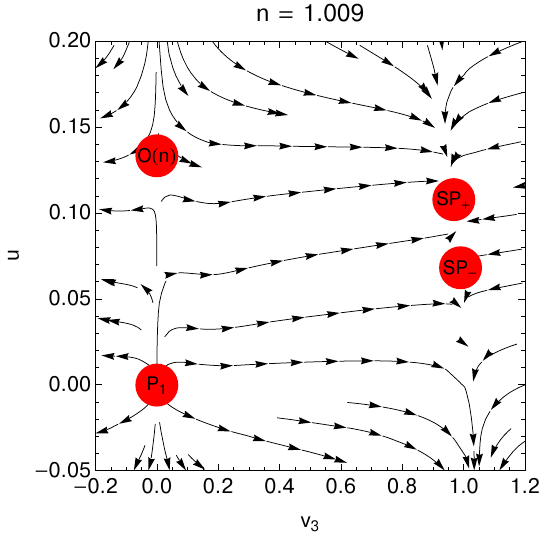}
    \includegraphics[scale=0.65]{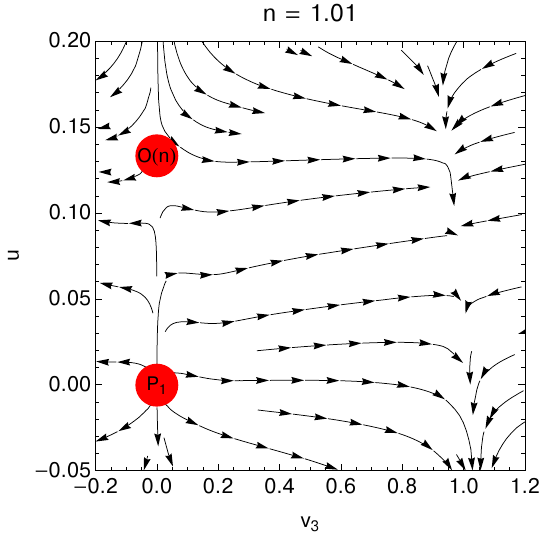}
    \caption{Renormalization group flow in $d=3.6$ in the plane $(v_3,u)$ for $n \in \{1,1.009,1.01\}$ and $k=4$. For the sake of simplicity, we have imposed $r=0$ and $\partial_t w_4=0$. The latter equation has two solutions and we choose the only one involving a  critical FP.}
    \label{fig:flowperturd36qc}
\end{figure}

For $n=1+\delta$ with $0<\delta\ll1$, the flow of $u$ depends on all the couplings and is therefore modified by a term proportional to $\delta$:

\begin{equation}
\label{flow-u-A(v)}
 \partial_t u=-\epsilon u+ 3 u^2 I_3+ A\,\delta+\mathcal{O}\big(\delta^2\big)+\mathcal{O}\big(\epsilon^3\big)+\mathcal{O}\big(\epsilon\delta\big).
\end{equation}
where $A$ depends on both isotropic and anisotropic couplings. From Eq.~\eqref{flow-u-A(v)}, we thus find two possible FPs that we call SP$_\pm$ and that correspond to:
\begin{equation}
 u_\pm^*=\frac{\epsilon\pm\sqrt{\epsilon^2-12 I_3 A_*\delta}}{6 I_3}.
 \label{SP-u}
\end{equation}
Let us assume that at the FP,  $A_*>0$ which is what we find at LPA and LPA'. Then, these FPs can only exist if
\begin{equation}
\label{realsolsSPpm}
 \epsilon^2\ge 12 I_3 A_*\delta,
\end{equation}
or, equivalently,
\begin{equation}
\label{perturbativqc}
q\leq q_c(d)=2+\frac{\epsilon^2}{12 I_3 A_*}+\mathcal{O}(\epsilon^3)\equiv 2+a \epsilon^2+\mathcal{O}(\epsilon^3).
\end{equation}
It is easy to compute the stability of these two FPs and we find that SP$_+$ is once unstable and is thus critical while SP$_-$ is twice unstable and is thus tricritical. This is consistent with the cartoon of the LPA' Renormalization Group flow shown in Fig.~\ref{fig:flowperturd36qc} and  with the fact that when $\delta\to 0$ at fixed $\epsilon$, SP$_+$ collides with the Wilson-Fisher FP of the Ising model and SP$_-$ with the Gaussian FP. 

As expected, the mechanism for switching from a first to a second order transition is the collision between two FPs, one being critical (SP$_+$) and the other one (SP$_-$) tricritical. This occurs when $u_+^*=u_-^*$ which determines the equation of the curve $q_c(d)$.

\begin{table}
	\caption{\label{T:O1} Anisotropic coupling constants and $\lim_{\delta\to0} u_6/\delta$ for $d=4$ and $q=2$ at SP$_{\pm}$. The coefficient $a$ defined in Eq.~(\ref{perturbativqc}) is computed at successive orders of the field truncation. The column LPA' represents an estimate of the nontruncated LPA' result obtained from the highest implemented order $k=9$ and the error is the difference with the previous order $k=8$. Notice that this error bar takes care only of the error coming from the field expansion and not from the truncation of the DE at LPA'.}
\begin{ruledtabular}
		\begin{tabular}{l|l l l l l l | l}
			order\,$k$ & 4 & 5 & 6 & 7 & 8 & 9 & LPA' \\
			\hline
			$v_3$ & 0.921 & 1.020  & 1.021  & 1.003 & 0.996 & 0.999 & 0.999(3) \\
			$w_4$ & 0.772 & 1.237 & 1.325 & 1.278 &  1.250 & 1.256 & 1.256(6) \\
			$v_5$ & & -1.027 & -1.520 & -1.543 & -1.484 & -1.482 & -1.482(2)  \\
            $x_5$ & & 1.143 & 1.664 & 1.640 & 1.556 & 1.560 & 1.560(6) \\
			$v_6$ & & & -1.584 & -2.266 & -2.339 & -2.308 & -2.31(3) \\
          $w_6$ & & & -2.483 & -3.087 &  -2.802 & -2.696 & -2.7(1) \\
			$y_6$  &  & & 1.294 & 1.524 & 1.309 & 1.253 & 1.25(6) \\
            $v_7$ & & & & 0.857 & 0.452 & -0.002 & 0.0(5) \\
            $w_7$  & & & & -7.462 & -9.819 & -9.651 & -9.7(1)  \\
             $x_7$ & & & & 	-0.971 & 0.873 & 2.196 & 2(1)\\
             $z_7$ & & & & -0.142 & -1.132 &	-1.678 & -1.7(6)\\
             $u_6/\delta$ & & & -0.809 & -1.336 & -1.104 & -1.105 & -1.105(1)\\
			\hline
   a & 0.053 & 0.084 & 0.113 & 0.098 & 0.106 & 0.104 & 0.104(2) 
		\end{tabular}
   \end{ruledtabular}
\end{table}
Up to now, the determination of $q_c(d)$ is exact in the infinitesimal neighborhood of $d=4$ and $q=2$, except for the value of $A_*$ that we have assumed to be positive. However, to calculate $a$ or $A_*$ an approximation going beyond perturbation theory must be performed. Here we use the LPA' to compute them. To do so requires to obtain the leading behavior of the right hand side of Eq. \eqref{floww4} in both $\epsilon$ and $\delta$. We have shown in Sec.~\ref{sssection:perturbFP} that for all the perturbative FPs in $\epsilon$, $v_3^*=0$ and all of them are multicritical. Therefore, the critical FP, if any, cannot be fully perturbative. We assume now, and this will be checked below, that this critical FP corresponds to $v_3^*\neq0$.
The exact analysis performed above shows that for  SP$_\pm$, $r,u \sim O(\epsilon)$, $u_{a\ge6}^*\sim\delta$ and $v^*, w^*, \dots\sim O(1)$.

Inserting these scalings in Eq. \eqref{floww4} and performing the double expansion in $\epsilon$ and $\delta$, we obtain:
\begin{align}
\label{A(v)-LPA}
A_{\rm LPA'}(v)&=-\frac{1}{5}v_6 I_2
+\Big(\frac{14}{5}v_3 v_5-\frac{4}{3}w_4^2\Big)I_3
+8 v_3^2 w_4 I_4-4 v_3^4 I_5\nonumber\\
&-\frac 1 2 \lim_{\delta\to0}\frac{u_6}{\delta} I_2.
\end{align}
It is important to notice that although the LPA is one-loop exact, the calculation of $A$ is not controlled by a one-loop analysis because it depends on anisotropic couplings that are not small near $d=4$.  To overcome this difficulty, we use here the LPA and LPA', see Sec.~\ref{ssection:floweqLPA}, which makes the calculation of $A$ in Eq.~\eqref{A(v)-LPA} approximate.  Another source of error in our calculation of $A$ comes from the field expansion that we have to implement when $n$ is not an integer. This error is however under control as can be seen in Table~\ref{T:O1} where it is manifest that the coupling constants of lowest orders involved in Eq. \eqref{A(v)-LPA} converge rather fast with the order of the field truncation.  It is also important to notice that since the couplings $v^*, w^*, \dots$ are of order 1, SP$_\pm$ are not fully perturbative FPs even with respect to the double expansion in $\epsilon$ and $\delta$ and we call them for this reason semi-perturbative FPs, hence their names SP$_\pm$.

We have computed the coefficient $a$ in Eq.~\eqref{perturbativqc} up to order 9 in the field expansion both by using Eqs. \eqref{perturbativqc} and \eqref{A(v)-LPA} in $d=4$ and by extrapolating the curve $q_c(d)$ to $d=4$. This curve is obtained as the location of the collision of SP$_+$ and SP$_-$ when $q$ is varied. When they collide, the first irrelevant eigenvalue of the linearized flows around these FPs vanishes. From a numerical point of view, we find it more convenient and accurate to characterize the curve $q_c(d)$ as the value of $q$ at fixed $d$ where this eigenvalue vanishes rather than looking for the value of $q$ where both FPs have disappeared. The two methods used to compute $A$, either by extrapolation of the curve $q_c(d)$ to $d=4$ and by using Eq.~\eqref{A(v)-LPA} in $d=4$ yield the same results up to numerical errors. This shows that the scalings in $\delta$ of the different coupling constants assumed to derive Eq. \eqref{A(v)-LPA} are indeed correct.

The constant $I_3 A_{\rm LPA}(v)$ in Eq.~\eqref{SP-u} is dimensionless and it is easy to check that it is therefore the same when expressed in terms of dimensionful or dimensionless quantities. This allows us to use the FP values of the couplings to compute  it. Moreover, all integrals involved in Eq.~\eqref{A(v)-LPA} can be computed in $d=4$ by taking $\eta=r=0$ and with the $\Theta-$regulator defined in Eq.~\eqref{reg-theta}, they are all equal to one.

We show in Table~\ref{T:O1} the values of the non-O($n$)-invariant couplings in $d=4$ and of the parameter $a$ defined in Eq.~\eqref{perturbativqc} for different orders $k$ of the field truncation. The evolution of $a$ with $k$ clearly indicates that this number converges to $0.104(2)$ where the error bar takes only into account the error induced by the field truncation and not the one coming from neglecting the higher orders of the derivative expansion.

Let us finally discuss the case $d>4$. For $q>2$ and imposing $u^*$ to be real again requires that (\ref{realsolsSPpm}) is fulfilled. However,  both $u_\pm^*$ are  negative. While the meaning of FP potentials for noninteger values of $q$ is not obvious, it is reasonable to assume that negative values of $u$ are unacceptable which would imply that the phase transition is of first order. On the other hand, if $\delta<0$ both $u_\pm^*$ are real and only $u_+^*$ is positive and the transition is of second order. We conclude that for $d>4$, the transition is probably of second order if and only if $q\leq 2$, as previously suggested in the literature \cite{Wu:1982ra,Newman:1984hy,BenAliZinati:2017vjy}.

\subsection{The critical line $q_c(d)$} 
\label{critical-qc-subsec}

In this subsection, we extend the analysis performed above around $d=4$ to lower dimensions. As discussed in the previous section, we use the LPA' and a field expansion truncated to order $k\le9$. The resulting equations are (\ref{flowr}) to (\ref{floww4}) for lower order couplings and  can be found in the supplementary material for  higher order couplings.

\begin{figure}[t!]
    \centering
    \includegraphics[scale=0.55]{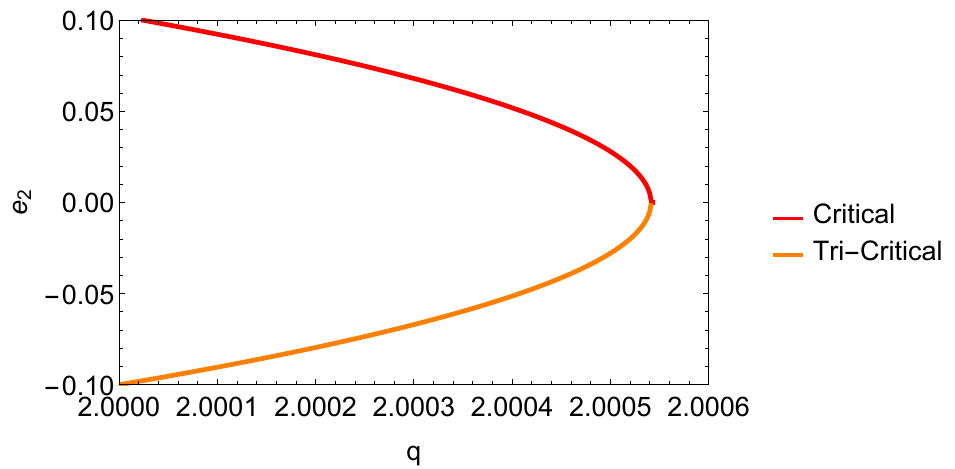}
    \caption{Second most relevant eigenvalue as a function of $q$ for the $k=4$ truncation in $d=3.9$ for the critical (red) and the tricritical (orange) FPs.}
    \label{fig:eigenplot}
\end{figure}

\begin{figure}[t!]
    \centering
    \includegraphics[scale=0.85]{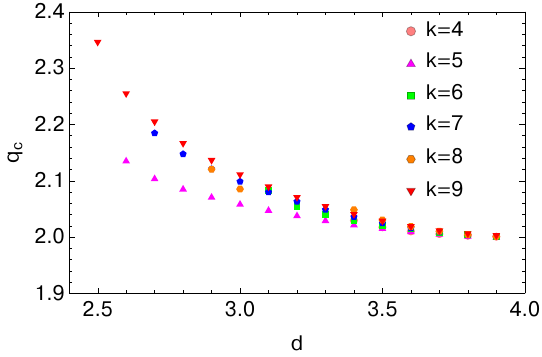}
    \caption{Curve $q_c(d)$ for each order $k$ of the field expansion in LPA'. The lower dimensions that can be reached for each order of the field expansion are: $d = 3.4$ for $k = 4$, $d = 2.6$ for $k =5$, $d = 3.1$ for $k = 6$, $d=2.7$ for $k=7$, $d=2.9$ for $k=8$ and $d=2.5$ for $k=9$. Notice that for $k=8$ no reliable determination of $q_c(d)$ can be obtained in the range $3.0<d<3.4$.} 
    \label{fig:qcvsdLPAp}
\end{figure}

\begin{figure}[t!]
    \centering
    \includegraphics[scale=0.85]{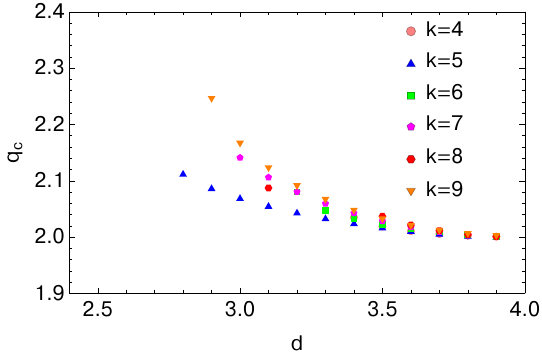}
    \caption{Curve $q_c(d)$ for each order $k$ of the field expansion in LPA. The lower dimensions that can be reached for each order of the field expansion are: $d = 3.4$ for $k = 4$, $d = 2.7 $ for $k =5$, $d = 3.2$ for $k = 6$, $d = 3.0$ for $k=7$, $d=3.1$ for $k=8$ and $d= 2.9$ for $k=9$. Notice that for $k=8$ no reliable determination of $q_c(d)$ can be obtained in the range $3.1<d<3.5$. } 
    \label{fig:qcvsdLPA}
\end{figure}
As said above, the curve $q_c(d)$ is the location in the $(d,q)$-plane where the critical and tricritical FPs collide. Equivalently, it is for each $d$, the value of $q$ for which both  the first irrelevant eigenvalue of the flow at the critical FP vanishes and the second most irrelevant eigenvalue of the flow at the tricritical FP vanishes. For $q>q_c(d)$, the two FPs have disappeared and, hence, the transition is of first order. More precisely, for $q>q_c(d)$, the two FPs are complex. An example of the variation of the second most relevant eigenvalue $e_2$ with $q$ is shown in Fig.~\ref{fig:eigenplot} for $d=3.9$ and $k = 4$. The curve $q_c(d)$ is given in Fig.~\ref{fig:qcvsdLPAp} for $k=4,\cdots, 9$.

At each order $k$ of the field expansion, we have numerically found that at sufficiently small values of $d$ the collision of FPs does no longer occur. \footnote{More precisely, we find that the curve $q_c(d)$ can either show at small $d$ an unphysical jump at a given order $k$ (e.g. for $d=3$ and $k=6$) or can vary so much from order $k$ to order $k+1$ that it is a clear indication of the nonconvergence of the field expansion, at least for the values of $k$ we are able to implement. We also observe that the convergence is better when we consider only the results obtained for odd values of $k$.} We consider that our calculation is no longer under control in these dimensions because of a lack of reliability caused by the field expansion. As expected, typically the dimension where the field expansion no longer works decreases when $k$ increases and the larger $k$ the better the convergence of the field expansion. Quite unexpectedly, we observe, without being able to explain it, that the convergence of the expansion is much better for odd values of $k$. For instance, even if $k=8$ yields results for $d\in[2.9,3]\cup[3.4,4]$ compatible with  those obtained both for $k=7$ and $k=9$, the mechanism of annihilation of FPs for $d\in[3,3.4]$ does not take place which is clearly an anomaly. This kind of anomaly does not occur for odd values of $k$ and moreover the dimension where the expansion does no longer work is systematically much smaller for odd $k$ than for even $k$.

Our results clearly show that our determination of $q_c(d)$ at the level of the LPA' is under control at least for $d\in[2.9,4]$. We find in particular at LPA': $q_c(3)=2.10$ for $k=7$ and $q_c(3)=2.11$ for $k=9$ and thus $q_c(3)=2.11(1)$. It is important to realize that the error bar given previously only takes  into account the error induced by the field truncation to order $k=9$ and not the error coming from truncating the DE at LPA', which produces a supplementary error.

A rough estimate of the error coming from the truncation of the DE at its lowest order is the difference between the determinations of $q_c(d)$  with either the LPA (see Fig.~\ref{fig:qcvsdLPA}) or the LPA'.
The rationale behind this choice is that the LPA' includes the RG evolution of $Z_k$ whereas the LPA does not. This is of course only an indication of the impact of the renormalization of the derivative terms on $q_c(d)$ and should not be taken as a precise value of the error bar. In particular, the $Z_k$ term that differentiates the LPA and LPA' is an isotropic ($O(n)-$invariant) term. Therefore, when $d$ approaches four, its contribution is suppressed for reasons discussed in Sect.~\ref{neard4}. This implies that for $d\to4^-$, the error coming from neglecting higher orders of the DE is underestimated for quantities that are sensitive to the anisotropic sector.

\begin{figure}[t!]
    \centering
    \includegraphics[scale=0.85]{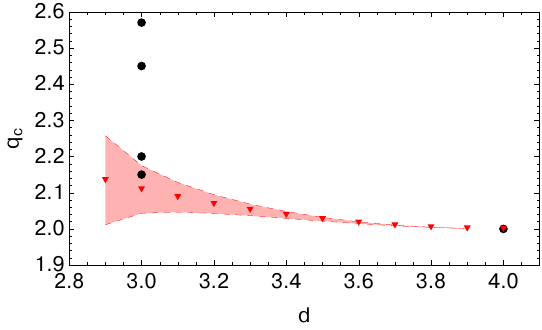}
    \caption{Curve $q_c(d)$ in LPA' and $k=9$ (red triangles). The red region represents an estimate of the confidence intervals of our results. The black points are previous results: $q_c(d = 4-\epsilon) = 2+\mathcal{O}(\epsilon^2)$\cite{Newman:1984hy} and
    $q_c(d=3)=2.15$\cite{GrollauTarjus2001}, $2.2$\cite{Nienhuis81}, $2.45$\cite{lee1991three} and $2.57$\cite{Kogut1982}.}
    \label{fig:qcvsdLPAvsLPAp}
\end{figure}

A general estimate of the errors generated by successive orders of DE has been proposed and tested successfully for both $O(N)$ models \cite{Balog2019,DePolsi2020,DePolsi2021} and for a model with $\mathbb{Z}_4$ anisotropies \cite{Chlebicki2022}. This requires to take into account 
at least the second order of the DE, which is beyond the scope of the present work. We can, however, expect that our rough error bar estimate is appropriate for quantities dominated by the isotropic sector, such as the exponents $\nu$ or $\eta$.

We show in Fig.~\ref{fig:qcvsdLPAvsLPAp} our final estimate of the curve $q_c(d)$. The central values are obtained for $k=9$ with LPA'. An estimate of the confidence intervals is represented by a red region obtained by summing the errors coming from the field expansion and the DE. For $k=9$, we find  $q_c^{\rm LPA}(d=3)-q_c^{\rm LPA'}(d=3)=0.06$ which is much larger than the error coming from the field truncation which is only $0.01$. Our final estimate is therefore $q_c(d=3)=2.11(7)$.

\begin{figure}[t!]
    \centering
    \includegraphics[scale=0.95]{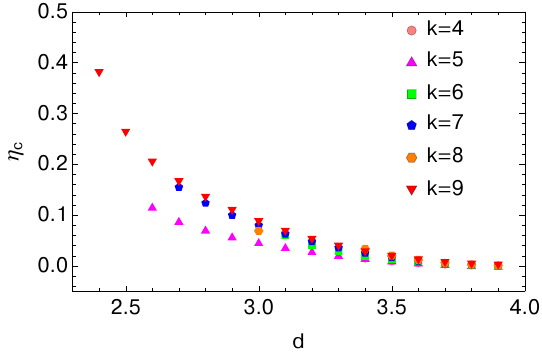}
    \caption{Anomalous dimension $\eta(q = q_c) = \eta_c$ as a function of $d$ for all implemented orders of the field expansion. These values of $\eta_c$ suggest that the LPA' is not sufficiently reliable to compute $q_c$ below $d<3$.}
    \label{fig:etacvsd}
\end{figure}

For  $d<3$, the difference between the LPA and the LPA' can be qualitative. In particular for $k=9$, the procedure to estimate the curve $q_c(d)$ does no longer work for the LPA below $d=2.9$ whereas the LPA' works down to $d=2.5$. For  $d\simeq 2.9$, we consider that our approximation scheme is no longer reliable even if $q_c(d)$ can be computed. Another indication of the limitations of our approximations at low dimension comes from the fact that the anomalous dimension along the curve $q_c(d)$ grows rapidly for dimensions $d<3$, as can be seen in Fig.~\ref{fig:etacvsd}.

Let us finally notice that since for $q=2$ and $q=3$ only the  invariants $\rho$ and $\tau_3$ play a role we could have naively  expected that they would go on playing a dominant role for all values in this range of $q$. This turns out to be wrong and for $k=9$ for instance it is quantitatively important to include all invariants up to $\tau_9$.

\section{Summary and outlook}

In the present paper, we  study the $q-$state Potts model for arbitrary real values of $q$ and $d$  at leading order of the Derivative Expansion (Local Potential Approximation) of the Non-Perturbative Renormalization Group.  We  also implement an improved variant commonly reffered to as the LPA' which allows for an anomalous dimension of the field. Our main goal is to compute the curve $q_c(d)$ which is the boundary between a first  and a second order phase transition region, respectively for $q>q_c(d)$ and $q<q_c(d)$. 

For $q\in\mathbb N$, the free energy associated with the Potts model depends on $q-1$ independent invariants under the permutation symmetries. On the other hand,  for non-integer $q$, it depends on infinitely many invariants. The implementation of the LPA and LPA' therefore requires an additional approximation, so as to deal with only a finite number of them. The field expansion of  the effective potential is such an approximation because it keeps only a finite number of these invariants when truncated to a finite order \cite{Newman:1984hy,BenAliZinati:2017vjy}. In a previous work \cite{Newman:1984hy}, such an expansion was implemented in the context of Wilson's RG up to the sixth power of the field. This did not allow the authors to go reliably to dimensions below $d\sim 3.4$. We extend this result with NPRG up to the ninth order which allows us to reach  $d=3$ in a controlled way. We obtain $q_c(d=3)=2.11(7)$, which is in line with previous studies and which confirms that the phase transition for $q=3$ in $d=3$ is of first order \cite{Kogut1982,lee1991three,janke1997three,GrollauTarjus2001,Chester:2022hzt}. 

The study of successive orders of the field expansion allows us to test its convergence. We observe that it deteriorates progressively as the dimension is decreased, as expected. In addition, the comparison between the LPA and the LPA' allows us to obtain a rough estimate of the influence of the renormalization of derivative terms and thus to analyze how reliable these approximations are.
This analysis shows that at this level of approximation our estimate of $q_c(d)$ is no longer under control below $d \lesssim 2.8$ and that it is neither  possible to reach $d=2$ in a controlled way nor the dimension where $q_c=3$.

A possible extension of the present work is to analyze the $q=0$ and $q=1$ cases  which were analyzed by similar methods in Ref.~\cite{BenAliZinati:2017vjy}. The most interesting dimension, $d=3$, was out of reach of this study that was performed at  order $k=6$ of the field expansion. It would be interesting to see whether the expansion up to order $k=9$ that we have implemented in the present work  allows us to reliably study the three dimensional case also for these values of $q$.

In addition to these physical applications, there are two different ways of going beyond the present analysis.

First, if we want to keep $q$ arbitrary, we have to deal with an infinite number of invariants which forces us to perform a field expansion. Even if we could imagine including second order DE terms, which is most probably extremely tedious, it is not at all clear that going on performing a field expansion would allow us reaching $d=2$ because of problems of the non-convergence of this expansion in low dimensions. A possible way out of this difficulty could be to work fully functionally with the isotropic  invariant $\rho$ and to expand in the anisotropic couplings (see, for example, \cite{Delamotte:2003dw,Chlebicki2022}). This intermediate procedure seems feasible and we plan to implement it in the near future.

Second, we can avoid the problem of the infinite number of invariants by considering only integer values of $q$ and, at first, $q=3$ where there are only two invariants, $\rho$ and $\tau_3$. As explained above, a reliable determination of $d_c(q=3)$ surely requires to implement the second order of the derivative expansion. We are currently analyzing the corresponding flow equations using the same techniques  developed in \cite{Chlebicki2022} for $\mathbb Z_4$-invariant systems and in \cite{Delamotte:2015zgf} for the study of frustrated magnetic systems. We expect that the implementation of the second-order  of the DE will enable us not only to reliably calculate $d_c(q=3)$ and study the two-dimensional case, but ideally also to determine error bars in dimensions where LPA' results are available.

\begin{acknowledgments}
We are very grateful to Alessandro Codello for valuable comments on the manuscript. C. S. and N. W. thank for the support of the Programa de Desarrollo de las Ciencias Básicas (PEDECIBA). This work received the support of the French-Uruguayan Institute of Physics project (IFU$\Phi$) and from the grant of number FCE-1-2021-1-166479 of the Agencia Nacional de Ingestigaci\'on e Innovaci\'on (Uruguay).
\end{acknowledgments}

\bibliographystyle{apsrev4-1}
\bibliography{Potts_q.bib}

\appendix

\section{Explicit construction of the vectors $\vec e^{\,(\alpha)}$}
\label{explicit vectors}

We  proceed iteratively to construct the vectors $\vec e^{\,(\alpha)}$. We first consider small values of $q$ and then generalize to arbitrary integers $q$.

\subsection{The $q=2$ case}

When $q=2$, corresponding to the Ising case, the number of components $n$ of the $\vec{e}^{(\alpha)}$ is one, that is, they are numbers:
\begin{equation}
 e^{(1)}=+1, \hspace{2cm} e^{(2)}=-1.
 \label{eqn:q2vectors}
\end{equation}
Here we choose the vectors to be of unit norm (as usually done in the Ising case).
In this particular case, this choice coincides with the normalization employed in \cite{BenAliZinati:2017vjy}.

\subsection{The $q=3$ case}

For $q=3$, that is, $n=2$, the basis vectors are planar. They join the center and the vertices of an equilateral triangle, see Fig.~\ref{fig:vectors}: 
\begin{equation}
 \vec e^{\,(1)}=(1,-\lambda_2/2), \hspace{.5cm} \vec e^{\,(2)}=(-1,-\lambda_2/2), \hspace{.5cm} \vec e^{\,(3)}=(0,\lambda_2).
\end{equation}
Note that the axes and normalizations are chosen such that the horizontal components of the first two vectors are identical to those of the $q=2$ model. We also impose the barycenter condition, Eq. (\ref{baricenter}). Moreover, $\lambda_2$ is fixed by imposing that the three
vectors have the same norm:
\begin{equation}
 \lambda_2^2=1+\frac{\lambda_2^2}{4},
\end{equation}
or, equivalently,
\begin{equation}
\label{lambda2}
 \lambda_2=\frac{2}{\sqrt{3}}.
\end{equation}
With this choice, their norm is neither 1 nor $\sqrt{n}$. The advantage of the present choice is that the first $n-1$ components of the first $n$ vectors $\vec e^{\,(\alpha)}$ are identical to those of the $(q-1)$-Potts models. As shown below, this implies that the projection to lower dimensional hyper-planes of the various tensors are identical for different values of $q$. This implies, in particular, the properties  given in Eqs. (\ref{projectionrho},\ref{projectiontau}). For other normalizations, this is only the case up to a normalization factor.

\subsection{The $q=4$ case}

One can generalize the previous construction to higher values of $q$. For the sake of clarity, we present now the $q=4$ case. 

For $q=4$, the vectors $\vec e^{\,(\alpha)}$ join  the barycenter of a regular tetrahedron to its vertices, as in Fig. \ref{fig:vectors}. As above, the first two components of the first three vectors are taken identical to those of the $q=3$ case and the third one is related to the fourth vector by imposing the barycenter condition Eq. (\ref{baricenter}):
\begin{align}
 &\vec e^{\,(1)}=(1,-1/\sqrt{3},-\lambda_3/3) \hspace{1cm} \vec e^{\,(2)}=(-1,-1/\sqrt{3},-\lambda_3/3) 
 \nonumber\\
&\vec e^{\,(3)}=(0,2/\sqrt{3},-\lambda_3/3) \hspace{1.2cm} \vec e^{\,(4)}=(0,0,\lambda_3).
\end{align}
As above, $\lambda_3$ is fixed by imposing that all vectors have equal norm, which yields: 
\begin{equation}
 \lambda_3^2=\lambda_2^2+\frac{\lambda_3^2}{9},
\end{equation}
or, equivalently,
\begin{equation}
 \lambda_3=\frac{\sqrt 3}{\sqrt{2}} .
\end{equation}

\subsection{The general case}

We can now build the $q$ basis vectors $\vec e^{\,(\alpha)}$ for general integer values of $q$.

The $q$ vectors $\vec e^{\,(\alpha)}_{q\, \rm state }$ join the barycenter of an $n$-dimensional hyper-tetrahedron to its vertices. The first $q-1$ vectors are chosen such that their first $q-1$ components are identical to those of the $q-1$ basis vectors $\vec e^{\,(\alpha)}_{q-1\, \rm state }$  and their last component is $-\lambda_n/n$. The last vector, $\vec e^{\,(q)}_{q\, \rm state }$, is chosen to be $(0,\cdots,0,\lambda_n)$. Thus, the $q\times q$ matrix $M^{(q)}$ of the components of the $q$ vectors $\vec e^{\,(\alpha)}_{q\, \rm state }$: $M^{(q)}_{\alpha\beta}=\left(\vec e^{\,(\alpha)}_{q\, \rm state }\right)_\beta$, is :
\begin{equation}
 M^{(q)}=
\left(\begin{matrix}
 \ M^{(q-1)}   & \begin{matrix} 0 \\ \vdots \\ 0\end{matrix} \\
    \begin{matrix}\displaystyle{ -\frac{\lambda_n}{n}} & \cdots& \displaystyle{ -\frac{\lambda_n}{n}} \end{matrix} & \lambda_n
\end{matrix}\right).
\end{equation}
With this choice, the barycenter condition  at order $q$:
\begin{equation}
\sum_{\alpha=1}^{q} \vec e_{q\, \rm state}^{(\alpha)}=0
\end{equation}
becomes a trivial consequence of the barycenter condition at order $q-1$.
The parameter $\lambda_n$ is finally determined by imposing that all vectors are of equal norm:
\begin{equation}
\label{eqn:lambdarelation}
 \lambda_n^2=\lambda_{n-1}^2+\frac{1}{n^2}\lambda_n^2,
\end{equation}
or, equivalently,
\begin{equation}
 \lambda_n=\lambda_{n-1} \frac{n}{\sqrt{n^2-1}}.
\end{equation}
Using Eq. \eqref{lambda2}, one finds:
\begin{equation}
\label{norm2}
 \lambda_n= \sqrt{\frac{2\,n}{n+1}}.
\end{equation}

\subsection{Some useful properties}
We derive below a useful property of the vectors $\vec e^{\,(\alpha)}$. It reads:
\begin{equation}
\label{eqn:sumprope}
 \sum_{\alpha=1}^{n+1} e_i^{(\alpha)}e_j^{(\alpha)}=\frac{n+1}{n} |\vec e^{\,(\alpha)}|^2 \delta_{ij},
\end{equation}
which is valid for any normalization of the vectors. The proof can be done by induction. It is obvious for $q=2$ and we assume that it is true  at order $q-1$. By computing the left hand side of Eq. \eqref{eqn:sumprope}, we find:
\begin{align}
 &\sum_{\alpha=1}^{n+1} (e_{q\,\rm state})_i^{(\alpha)}(e_{q\,\rm state})_j^{(\alpha)}=\sum_{\alpha=1}^{n}\Big((e_{q-1\,\rm state})_i^{(\alpha)}-\frac{1}{n} (e_{q\,\rm state})_i^{(q)}\Big)\nonumber\\
 &\times \Big((e_{q-1\,\rm state})_j^{(\alpha)}-\frac{1}{n} (e_{q\,\rm state})_j^{(q)}\Big)+ (e_{q\,\rm state})_i^{(q)}(e_{q\,\rm state})_j^{(q)}\nonumber\\
&= \delta_{ij} \big(\delta_{qi}-1)\big(\delta_{qj}-1) \lambda_{q-1}^2 \frac{n}{n-1}
-\frac{1}{n} (e_{q\,\rm state})_i^{(q)}\sum_{\alpha=1}^{n} (e_{q-1\,\rm state})_j^{(\alpha)}\nonumber\\
&-\frac{1}{n} (e_{q\,\rm state})_j^{(q)}\sum_{\alpha=1}^{n} (e_{q-1\,\rm state})_i^{(\alpha)}+\delta_{iq}\delta_{jq} \lambda_q^2 \Big(1+\frac{1}{n}\Big)\nonumber\\
&=\frac{n+1}{n} |e_{q\,\rm state}^{(\alpha)}|^2 \delta_{ij}.
\end{align}

Using the normalisation Eq.~\eqref{norm2}, we find:
\begin{equation}
\label{T2tensor}
 \sum_{\alpha=1}^{n+1} e_i^{(\alpha)}e_j^{(\alpha)}=2 \delta_{ij}.
\end{equation}
This is one example of the simplifications that comes from this choice of normalization.

It is important to note that most properties of the tensors can be deduced from the  properties Eqs. \eqref{baricenter}, (\ref{T2tensor}) and (\ref{scalarprod}) that we summarize here:
\begin{align}
 \label{usefulid}
 &\sum_{\alpha=1}^{q} \vec e^{\,(\alpha)}=0, \nonumber\\
 &\sum_{\alpha=1}^{n+1} e_i^{(\alpha)}e_j^{(\alpha)}=2 \delta_{ij},\\
& \vec e^{\,(\alpha)}\cdot \vec e^{\,(\beta)}=2
 \Big( \delta_{\alpha\beta}-\frac{1}{n+1}\Big).\nonumber
\end{align}
They can be naturally extended to non-integer values of $q$, as done in \cite{BenAliZinati:2017vjy} with another normalization condition. 

\section{Properties of the tensors $\bar{T}^{(p)}$}
\label{tensorprops}
First, the tensors $\bar{T}^{(p)}$ are completely symmetric. 
Second,  using Eq. \eqref{usefulid} and the explicit construction of the vectors $\vec e^{\,(\alpha)}$ given above, we find: 
\begin{equation}
\label{eqn:minik}
 \bar{T}^{(p)}_{i_1 i_2\dots i_p}=\frac 1 2\sum_{\alpha=1}^{q} e^{\,(\alpha)}_{i_1}e^{\,(\alpha)}_{i_2}\dots e^{(\alpha)}_{i_p}
 =\frac 1 2\sum_{\alpha=1}^{\min\{i_k\}+1} e^{(\alpha)}_{i_1}e^{(\alpha)}_{i_2}\dots e^{(\alpha)}_{i_p}.
\end{equation}
Third, as a consequence, if at least one index is 1:
\begin{equation}
\label{eq:minik2}
 \bar{T}^{(p)}_{1 i_2\dots i_p}=\frac 1 2\sum_{\alpha=1}^{q} e^{(\alpha)}_{1}e^{(\alpha)}_{i_2}\dots e^{(\alpha)}_{i_p}
 =\frac 1 2\sum_{\alpha=1}^{2} e^{(\alpha)}_{1} e^{(\alpha)}_{i_2}\dots e^{(\alpha)}_{i_p}.
\end{equation}
It follows that if the number $k$ of indices equal to 1 in $ \bar{T}^{(p)}$ is odd, it vanishes since: 
\begin{equation}
 \bar{T}^{(p)}_{11\dots 1 i_{k+1}\dots i_p}=\frac 1 2\sum_{\alpha=1}^{q} (e^{(\alpha)}_{1})^k e^{(\alpha)}_{i_{k+1}}\dots e^{(\alpha)}_{i_p}
 =\frac 1 2 e^{(1)}_{i_{k+1}}\dots e^{(1)}_{i_p}\sum_{\alpha=1}^{2} (e^{(\alpha)}_{1})^k,
\end{equation}
which is zero if $k$ is odd.

Fourth, using Eq. (\ref{usefulid}), we can now prove Eq. (\ref{contractionid}):
\begin{align}
\bar{T}^{(p)}_{i_1i_2\dots i_{p-1}k}&\bar{T}^{(p')}_{j_1j_2\dots j_{p'-1}k}=\frac{1}{4}\sum_{\alpha,\beta}e_{i_1}^{(\alpha)}\dots e_{i_{p-1}}^{(\alpha)}e_{k}^{(\alpha)}e_{j_1}^{(\beta)}\dots e_{j_{p'-1}}^{(\beta)}e_k^{(\beta)} \nonumber\\
&=\frac{1}{2}\Big(\sum_{\alpha}
e_{i_1}^{(\alpha)}\dots e_{i_{p-1}}^{(\alpha)}e_{j_1}^{(\alpha)}\dots e_{j_{p'-1}}^{(\alpha)}\nonumber\\
&-\frac{1}{n+1}\sum_{\alpha,\beta}
e_{i_1}^{(\alpha)}\dots e_{i_{p-1}}^{(\alpha)}e_{j_1}^{(\beta)}\dots e_{j_{p'-1}}^{(\beta)}\Big)\nonumber\\
&=\bar{T}^{(p+p'-2)}_{i_1i_2\dots i_{p-1}j_1j_2\dots j_{p'-1}}-\frac{2}{n+1}\bar{T}^{(p-1)}_{i_1i_2\dots i_{p-1}}\bar{T}^{(p'-1)}_{j_1j_2\dots j_{p'-1}}.
\end{align}

In a similar way, one can prove a trace identity:
\begin{align}
\label{traceappendix}
\bar{T}^{(p)}_{i_1i_2\dots i_{p-2}jj}&=\frac{1}{2}\sum_{\alpha}e_{i_1}^{(\alpha)}e_{i_2}^{(\alpha)}\dots e_{i_{p-2}}^{(\alpha)}e_j^{(\alpha)}e_j^{(\alpha)}\nonumber\\
&=\frac{1}{2}\sum_{\alpha}e_{i_1}^{(\alpha)}e_{i_2}^{(\alpha)}\dots e_{i_{p-2}}^{(\alpha)}\frac{2n}{n+1}\nonumber\\
&=\frac{2n}{n+1}\bar{T}^{(p-2)}_{i_1i_2\dots i_{p-2}}.
\end{align}

\subsection{Some explicit tensor components}
\label{explicitT}
The properties presented before allows the explicit calculation of
several tensor elements. For example, for the tensor $T^{(3)}$:
\begin{align}
T^{(3)}_{112}&=\frac 1 2\sum_{\alpha=1}^{n+1} (e^{(\alpha)}_{1})^2 e^{(\alpha)}_{2} =\frac 1 2\sum_{\alpha=1}^{2} (e^{(\alpha)}_{1})^2 e^{(\alpha)}_{2}\nonumber\\
&=\frac 1 2\sum_{\alpha=1}^{2} e^{(\alpha)}_{2}=-\frac{\lambda_2}{2}=-\frac{1}{\sqrt{3}},
\end{align}
and
\begin{align}
T^{(3)}_{222}&=\frac 1 2\sum_{\alpha=1}^{n+1} (e^{(\alpha)}_{2})^3 =\frac 1 2\sum_{\alpha=1}^{3} (e^{(\alpha)}_{2})^3\nonumber\\
&=\frac 1 2\Big(-\frac{\lambda_2^3}{8}-\frac{\lambda_2^3}{8}+\lambda_2^3\Big)=\frac{3}{8}\lambda_2^3=\frac{1}{\sqrt{3}}.
\end{align}
 For the improved tensor $T^{(4)}$, we find:
\begin{align}
T^{(4)}_{1111}&=\frac 1 2\sum_{\alpha=1}^{n+1} (e^{(\alpha)}_{1})^4 -\frac{6n}{(n+1)(n+2)}\nonumber\\
&=\frac 1 2\sum_{\alpha=1}^{2} (e^{(\alpha)}_{1})^4 -\frac{6n}{(n+1)(n+2)}\nonumber\\
&=\frac{(n-1)(n-2)}{(n+1)(n+2)},
\end{align}
\begin{align}
T^{(4)}_{1122}&=\frac 1 2\sum_{\alpha=1}^{n+1} (e^{(\alpha)}_{1})^2 (e^{(\alpha)}_{2})^2 -\frac{2n}{(n+1)(n+2)}\nonumber\\
&=\frac 1 2\sum_{\alpha=1}^{2} (e^{(\alpha)}_{1})^2 (e^{(\alpha)}_{2})^2 -\frac{2n}{(n+1)(n+2)}\nonumber\\
&=\frac 1 3-\frac{2n}{(n+1)(n+2)}=\frac{(n-1)(n-2)}{3(n+1)(n+2)},
\end{align}
and 
\begin{align}
T^{(4)}_{2222}&=\frac 1 2\sum_{\alpha=1}^{n+1} (e^{(\alpha)}_{2})^4 -\frac{6n}{(n+1)(n+2)}\nonumber\\
&=\frac 1 2\sum_{\alpha=1}^{3} (e^{(\alpha)}_{2})^4 -\frac{6n}{(n+1)(n+2)}\nonumber\\
&=\frac{(n-1)(n-2)}{(n+1)(n+2)},
\end{align}

\section{Perturbative fixed point in $d=4-\epsilon$}
\label{perturbativefp}
As pointed out in Sect.~\ref{neard4} for $d=4-\epsilon$ there are four FPs that can be controlled  perturbatively. In this appendix, we analyze them at leading order in $\epsilon$. The first thing to note, as already mentioned in Sect.~\ref{neard4}, is that these FPs have an extra $\mathbb{Z}_2$ symmetry corresponding to inverting all components of the field ($\vec \phi \to -\vec \phi$) in addition to the ${\mathcal S}_q$ symmetry. This is due to the fact that for $d=4-\epsilon$, the FP value of the coupling constant $v_3$ vanishes ($v_3^*=0$). 

To study the four most relevant operators at the four perturbative FPs, it suffices (i) to use the equations (\ref{flowr}--\ref{floww4}) that to this order in $\epsilon$, become exact, (ii) to take into account that $\eta=\mathcal{O}(\epsilon^2)$ and (iii) that the constants corresponding to operators with more than four fields are of order $\epsilon^3$ or higher. This yields:
\begin{align}
\label{eq:flowpert}
 \partial_t r&=-2 r-\frac{(n+2)}{6}u I_2(r), \nonumber\\
 \partial_t v_3&=-\frac{2+\epsilon}{2} v_3+\Big(2u+\frac{6 (n-2)}{n+2} w_4 \Big)v_3 I_3(r), \nonumber\\
\partial_t u&=-\epsilon u+\Big(\frac{1}{3} (n+8) u^2+\frac{24 (n-2) (n-1)}{(n+1) (n+2)^2}w_4^2\Big)I_3(r),\nonumber\\
\partial_t w_4&=
-\epsilon w_4 +\Big(6 \frac{(n-1)(n-2)}{(1+n)(2+n)} w_4^2+4 u w_4\Big) I_3(r).
\end{align}
 For the linear perturbations around the different FPs, it is sufficient to consider $v_3$ at linear order in all flow equations since $v_3^*=0$. In practice, this implies to neglect completely $v_3$ in all equations except in its own flow equation.

In Eqs. \eqref{eq:flowpert}, we made explicit that the functions $I_n$ depend on $r$. Now, the equations clearly show that the FP values of $u$ and $w_4$ are of order $\epsilon$. Since the FP value of $r$ is of order $u$, it is also of order $\epsilon$. Thus, for finding the FPs at this order in $\epsilon$, it is sufficient to evaluate the functions $I_n(r)$ at $r=0$.
The dependence on $r$ must be kept, as usual, only when studying the linear perturbation in the coupling $r$ around each FP. In this way, we determine the FPs detailed below and their scaling exponents.

\subsection{Gaussian fixed point}

The Gaussian FP corresponds to $r^*=v_3^*=u^*=w_4^*=0$. In this case, the relevance of the linear perturbations can be determined by  dimensional analysis. It has two $O(n)-$invariant relevant directions. One has scaling dimensions $-2$ (corresponding to an exponent $\nu=1/2$) and the other has scaling dimension $-\epsilon$. In addition to these two relevant directions, it has two other ${\mathcal S}_q-$symmetric relevant directions: one with scaling dimensions $-1 -\epsilon/2$ which is odd under $\mathbb{Z}_2$ transformations and one with scaling dimension $-\epsilon$ and which is even under $\mathbb{Z}_2$ transformations.

\subsection{The $O(n)-$invariant fixed point}

The second FP is the $O(n)-$invariant Wilson-Fisher FP. Its coordinates are:
\begin{align}
  r^{O(n),*}&=- \epsilon \frac{(n+2) I_2}{4 (n+8) I_3}  \nonumber\\
  v_3^{O(n),*}&=0,\nonumber\\
  u^{O(n),*}&=3 \frac{\epsilon}{(n+8) I_3}\nonumber\\
  w_4^{O(n),*}&=0.
\end{align}
Among the four most relevant perturbations around this FP  two that are $O(n)-$invariant. One is the usual most relevant one for $O(n)$ models. It has scaling dimension $-2 + \frac{n+2}{n+8} \epsilon$ that corresponds to the exponent $\nu=1/2 + \frac{n+2}{4 (n+8)} \epsilon$). The other $O(n)-$invariant scaling perturbation is irrelevant, has scaling dimension $\epsilon$ and corresponds to the usual exponent  $\omega$ of $O(n)$ models. The two other perturbations are only  ${\mathcal S}_q-$invariant. One is relevant, with scaling dimension $-1-(n-4) \epsilon/(2(n+8))$ and is odd under $\mathbb{Z}_2$ transformations. The other is even under $\mathbb{Z}_2$ transformations and has dimension scaling $(4-n)\epsilon/(n+8)$. As a consequence, it is irrelevant for $n<4$ but becomes relevant if $n>4$, that is, the  $O(n)-$invariant FP is tricritical if $n<4$ and is tetracritical if $n>4$.

\subsection{First anisotropic perturbative fixed point}

The third perturbative FP is not $O(n)-$invariant but only ${\mathcal S}_q-$invariant. Its coordinates are:
\begin{align}
  r^{AP1,*}&=-\frac{ I_2(n-2)(n-1)}{12 I_3 (n^2-5n+8)} \epsilon  \nonumber\\
  v_3^{AP1,*}&=0,\nonumber\\
  u^{AP1,*}&=\frac{\epsilon (n-2)(n-1)}{I_3 (n+2)(n^2-5n+8)}\nonumber\\
  w_4^{AP1,*}&=\frac{\epsilon (n-4)(n+1) }{6 I_3 (n^2-5n+8)}.
\end{align}

The most relevant scaling operator around this FP is even under $\mathbb{Z}_2$ transformations and has scaling dimension $-2 + \frac{(n-2)(n-1)}{3 (n^2-5n+8)}\epsilon$ (corresponding to an exponent $\nu=1/2 + \frac{(n-2)(n-1)}{12 (n^2-5n+8)}\epsilon$). The second most relevant scaling operator is odd under $\mathbb{Z}_2$ transformations and has scaling dimension $-1 + \frac{(n-4) (n-1)}{2 (n^2-5n+8)} \epsilon$.
In addition to these two relevant operators, there are two operators that are even under $\mathbb{Z}_2$ transformations.
One of them is always irrelevant, having a scaling dimension $\epsilon$ for any $n$. The last scaling operator among these four has scaling dimension $\frac{(5-n)(n-4)}{3 (n^2-5n+8)}\epsilon$. As a consequence, it is relevant for $4<n<5$ and irrelevant (or marginal) for other values of $n$. This means that this FP is tricritical for $4<n<5$ and tetracritical otherwise.

\subsection{Second anisotropic perturbative fixed point}

As the previous one, the fourth and last perturbative FP  is not $O(n)-$invariant but only ${\mathcal S}_q-$invariant. Its coordinates are:
\begin{align}
  r^{AP2,*}&=-\frac{\epsilon (n+1)}{6 I_3 (n+3)} I_2 \nonumber\\
  v_3^{AP2,*}&=0,\nonumber\\
  u^{AP2,*}&=\frac{2 (n+1)}{I_3 (n+2)(n+3)}\epsilon \nonumber\\
  w_4^{AP2,*}&=\frac{\epsilon (n+1) }{6 I_3 (n+3)} .
\end{align}

The most relevant scaling operator around this FP is, as in the previous one, even under $\mathbb{Z}_2$ transformations and has scaling dimension $-2 +\frac{2(n+1)}{3 (n+3)}\epsilon$ (corresponding to an exponent $\nu=1/2 + \frac{(n+1)}{6 (n+3)}\epsilon$). The second most relevant scaling operator is, again, odd under $\mathbb{Z}_2$ transformations and has scaling dimension $-1 + \frac{(n-1)}{2 (n+3)} \epsilon$.
In addition to these two relevant operators, there are two others that are even under $\mathbb{Z}_2$ transformations.
One of them is, as before, always irrelevant, having a scaling dimension $\epsilon$ for any $n$. The last scaling operator among these four has scaling dimension $\frac{(n-5)}{3 (n+3)}\epsilon $. As a consequence, it is relevant for $n>5$ and irrelevant (or marginal) for other values of $n$. This means that this FP is tricritical for $n>5$ and tetracritical otherwise.

\end{document}